\documentclass[9pt]{article}

\usepackage{amsmath,amssymb}
\usepackage{microtype}
\usepackage{graphicx}
\usepackage{algorithm,algpseudocode}
\usepackage{mathrsfs}
\usepackage{epstopdf}
\usepackage{float}
\usepackage{url}
\usepackage{relsize}
\usepackage{subfloat}
\usepackage{subcaption}
\usepackage{caption}
\usepackage{multirow}
\usepackage{mathptmx}

\usepackage{amsbsy}
\usepackage{bm}
\usepackage{blkarray}
\newcommand{\Tr}{\mathrm{Tr}}

\newcommand{\R}{\bm{R}}
\newcommand{\G}{\bm{G}}
\newcommand{\x}{\bm{x}}

\newcommand{\q}{\bm{q}}

\begin{document}

\title{Integrating NOE and RDC using sum-of-squares relaxation for protein structure determination%\thanks{Grants or other notes
%about the article that should go on the front page should be
%placed here. General acknowledgments should be placed at the end of the article.}
}

\author{
Y. Khoo\thanks{Department of Physics, Princeton University, Princeton, NJ 08540,USA ({\tt ykhoo@princeton.edu}).}
\and
A. Singer\thanks{Department of Mathematics and PACM, Princeton University, Princeton, NJ 08544, USA ({\tt amits@math.princeton.edu}).}
\and
D. Cowburn\thanks{Department of Biochemistry, Albert Einstein College of Medicine, Bronx, NY 10461, USA ({\tt david.cowburn@einstein.yu.edu}).}
}

% The correct dates will be entered by the editor

\maketitle

\begin{abstract}
We revisit the problem of protein structure determination from geometrical restraints from NMR, using convex optimization.
It is well-known that the NP-hard distance geometry problem of determining atomic positions from pairwise distance restraints can be relaxed into a convex semidefinite program (SDP).
However, often the NOE distance restraints are too imprecise and sparse for accurate structure determination.
Residual dipolar coupling (RDC) measurements provide additional geometric information on the angles between atom-pair directions and axes of the principal-axis-frame.
The optimization problem involving RDC is highly non-convex and requires a good initialization even within the simulated annealing framework.
In this paper, we model the protein backbone as an articulated structure composed of rigid units. Determining the rotation of each rigid unit gives the full protein structure. We propose solving the non-convex optimization problems using the \emph{sum-of-squares} (SOS) hierarchy, a hierarchy of convex relaxations with increasing complexity and approximation power. Unlike classical global optimization approaches, SOS optimization returns a certificate of optimality if the global optimum is found. Based on the SOS method, we proposed two algorithms - RDC-SOS and RDC-NOE-SOS, that have polynomial time complexity in the number of amino-acid residues and run efficiently on a standard desktop. In many instances, the proposed methods exactly recover the solution to the original non-convex optimization problem. To the best of our knowledge this is the first time SOS relaxation is introduced to solve non-convex optimization problems in structural biology.

We further introduce a statistical tool, the Cram\'er-Rao bound (CRB), to provide an information theoretic bound on the highest resolution one can hope to achieve when determining protein structure from noisy measurements using any methodology.
Our simulation results show that when the RDC measurements are corrupted by Gaussian noise of realistic variance, both SOS based algorithms attain the CRB.

We successfully apply our method in a divide-and-conquer fashion to determine the structure of ubiquitin from experimental NOE and RDC measurements obtained in two alignment media, achieving more accurate and faster reconstructions compared to the current state of the art.
\end{abstract}

\section{Introduction}

The problem of positioning a set of points from geometrical constraints between them arises naturally when calculating the protein structure from Nuclear Magnetic Resonance (NMR) spectroscopy data. The best established structural calculation methods are based on the through-space dipole interaction of the Nuclear Overhauser Effect (NOE) \cite{kumar1980two,wuthrich2003nmr}.
The NOE gives rise to qualitative distance constraints of the following form
\begin{equation}
\label{distance constraint}
d_{nm}^\mathrm{lower} \leq \|\bm{x}_n - \bm{x}_m\|_2 \leq d_{nm}^\mathrm{upper}
\end{equation}
where $\bm{x}_n, \bm{x}_m$ are the coordinate positions of atoms $n$ and $m$, and $d_{nm}^\mathrm{lower}, d_{nm}^\mathrm{upper}$ are lower and upper bounds, respectively, for the Euclidean distance between these atoms. Since the NOE interaction between a pair of atoms scales as $ r^{-6}$, constraints for pairs of atoms that are more than 6 \AA  \ apart are too small and imprecise for use. For large molecules, the extraction of NOE restraints through resonance assignment is difficult and often leads to missing, ambiguous, or incorrect NOE distance measurements. Hence the inverse problem of positioning from distance constraints alone, also known as the distance geometry problem, can be challenging and even ill-posed \cite{xu2006new}. While multiple ingenious and interesting methods are used to address these issues \cite{mareuil2015improved,schmidt2012new}, obtaining a fully automated structural determination software based on NOE alone remains challenging. As noted in \cite{mareuil2015improved}, the process of filtering out the wrong NOE restraints may require manual intervention.

Residual dipolar coupling (RDC) measurements provide additional geometrical information involving pairs of atoms \cite{tolman1995rdc,tjandra1997direct}. RDC can be measured when the molecule ensemble in solution exhibits partial alignment with the magnetic field in an NMR experiment.
The RDC measurements have relatively high precision due to the slower $1/r^3$ decay of interaction, and provide alignment information involving pairs of atoms and the magnetic field. Under some technical assumptions, the RDC measurement $r_{nm}$ for atoms $n$ and $m$ is related to their positions in the following way:
\begin{equation}
\label{rdc constraint}
r_{nm} = \frac{(\bm{x}_n - \bm{x}_m)^T \bm{S} (\bm{x}_n - \bm{x}_m)}{d_{nm}^2},
\end{equation}
where $d_{nm} = \|\bm{x}_n - \bm{x}_m \|_2$ is the distance between atoms $n$ and $m$, and $\bm{S}$ is a $3\times 3$ symmetric matrix with vanishing trace, known as the Saupe alignment tensor
\cite{saupe1963high}. 
Roughly speaking, the eigenvectors of the Saupe tensor encode how the molecule aligns with respect to the magnetic field. Performing NMR experiments at different alignment conditions may lead to different Saupe tensors, and consequently different RDC measurements. While in principle both the Saupe tensor and the molecular structure are unknown, in this paper we assume that $S$ can be estimated a-priori \cite{losonczi1999order,zweckstetter2008nmr} and our goal is to determine the atom positions given $S$. We primarily focus on protein backbone structure determination from RDC data. For a detailed exposition of RDC and the Saupe tensor, we refer readers to the appendix and to \cite{lipsitz2004rdc,blackledge2005survey,tolman2006rdc}.
% iinsert citation abouve 1.	Saupe AZ. Kernresonanzen in kristallinen Flüssigkeiten und in kristallinflüssigen Lösungen. Teil I. Naturforsch. 1964;19a:161-71 

\subsection{Existing Approaches}
Most approaches to the structural determination problem apply a global optimization technique \cite{more1999distance,liwo1999protein,schwieters2003xplor,joo2015protein,joo2016contact} to obtain the global minima of a non-convex ``energy" function. The energy function includes pseudo-potential terms that restrain the pairwise interatomic distances (NOE), dihedral angles ($J$-coupling), packing (van der Waals interactions), and orientation with respect to a global magnetic field (RDC).

The mainstream approach to minimize the energy function is based on simulated annealing \cite{kirkpatrick1983optimization,guntert1997torsion,clore1998direct,schwieters2003xplor}. In simulated annealing, the ``tunneling'' mechanism pushes the solution out of a local minimum with a certain probability and the procedure can be run for many iterations in order to increase the chances of escaping local minima. In principle, this gives simulated annealing the versatility to deal with arbitrary non-convex energy functions, in particular, one can consider the following non-convex RDC potential term:
\begin{equation}
\bigg(r_{nm} - \frac{(\bm{x}_n - \bm{x}_m)^T \bm{S} (\bm{x}_n - \bm{x}_m)}{d_{nm}^2}\bigg)^2
\end{equation}
This RDC potential term yields, however, a rugged energy landscape with sharp local minima that hinders the success of finding the correct conformation in the absence of a good initial structure \cite{chen2012rdc,bax2001rdc}. For example, \cite{mukhopadhyay2014dynafold} reports that direct minimization of the RDC potential using simulated annealing can yield structures that are as much as 20 \AA\ away from the correct structure.
A popular way to initialize simulated annealing for protein structure determination from  RDC is by the molecular fragment replacement (MFR) approach \cite{kontaxis2005molecular}. MFR finds homologous short fragments of the protein in the Protein Data Bank with the aid of RDC and chemical shifts. The fragments are then merged together to form an initial structure to be locally refined by simulated annealing. However, using existing structures as initialization leads to model bias.  Moreover, there is still no guarantee that the initialization is good enough to avoid trapping at a local minimum.

Besides stochastic optimization,  a number of deterministic approaches based on branch and prune \cite{yershova2011rdc,cassioli2015bp} and dynamic programming \cite{mukhopadhyay2014dynafold} have been proposed more recently  to find the globally optimal backbone structure. In particular, RDC-ANALYTIC \cite{wang2004exact,wang2006polynomial,yershova2011rdc} exploits that in the presence of two RDC measurements per amino-acid, the torsion angles that determine the orientation of an amino-acid have $16$ possible value sets, and a solution tree with a total of $16^M$ possible structures can be built sequentially for a protein with $M$ amino-acids. The main advantage of branch and prune type methods is their ability to deal with sparse RDC datasets when used with an efficient adjunct pruning device such as the Ramachandran plot \cite{ramachandran1963stereochemistry} and NOE.
%It can also return multiple low-energy solutions when the protein has certain flexibility \cite{tripathy2012protein}. 
In addition, it can produce multiple low-energy solutions reflecting intrinisic flexibity \cite{tripathy2012protein}. 
Another approach with a similar flavor to the tree-searching based methods, REDCRAFT \cite{bryson2008redcraft}, performs Monte-Carlo sampling of the torsion angles of a protein based on the Ramachandran distribution. RDC measurements are then used to select the possible torsion angles. In general, the methods based on building the conformation space and pruning the unwanted conformations can lead to slow running times. Both REDCRAFT and RDC-ANALYTIC need an hour or two to solve for the structure of a typical size protein. Another approach with a different flavor is the dynamic programming approach in \cite{mukhopadhyay2014dynafold}. By casting the protein structuring problem from RDC as a shortest path problem, a solution can be obtained optimally and efficiently. However, it cannot readily incorporate additional information such as distance restraints to improve the solution quality.

A separate line of research is based on convex relaxation, in which the non-convex domain of an optimization problem is replaced by a convex domain. When the global optimum of the convex surrogate problem lies in the original domain, we can be sure that the original problem is solved. Otherwise, a rounding scheme can be used to project the solution from the convex set back to the original domain. For the distance geometry problem, semidefinite programming (SDP) relaxations \cite{so2007snl,biswas2006snlsdp,ding2010sensor} have been proposed. Under certain conditions on the distance measurements, it is shown that the solution to the NP-hard \cite{saxe1980embeddability} distance geometry problem can be computed in polynomial time \cite{so2007snl}. Since the introduction of the SDP relaxation, numerous efforts have been made for its computational speedup using additional relaxation \cite{wang2008esdp}, divide-and-conquer procedures  \cite{leung2009sdp,cucuringu2012eigenvector}, and facial reduction \cite{alipanahi2013determining}. While these methods are highly accurate in the presence of abundant distance restraints and do not suffer from local minima issues, their performance is unsatisfactory when lacking sufficient NOE measuremets (especially for large proteins due to spin diffusion \cite{prestegard2014sparse}). In such cases, it is crucial to refine the solution obtained by SDP relaxation by minimizing the original non-convex energy using another method such as simulated annealing.

\subsection{Overall approach}
We limit our attention to the calculation of protein backbone structure, leveraging the RDC and NOE measurements for the backbone. Unlike previous convex relaxation approaches that focused solely on distance constraints, we propose convex relaxations for backbone structure determination that simultaneously incorporate both NOE and RDC measurements. An additional advantage of this combination method is that it can provide accurate solutions even when using RDC alone.

We believe our proposed algorithm provides a solution to the Open Problem posed in \cite[Chapter~36]{donald2011algorithms}: {\em ``Use SDP and the concept of distance geometry with angle restraints to model RDC-based structure determination.''} In some sense, the structural calculation problem from RDC measurements of the form (\ref{rdc constraint}) can be regarded as the distance geometry problem in an inner product space (corresponding to the Saupe tensor) different from the standard Euclidean space.
Since the convex relaxations in \cite{so2007snl,biswas2006snlsdp} proposed for the distance geometry problem only involve the Gram matrix (inner product matrix) \cite{havel1998distance} of the atom coordinates in the Euclidean space, these methods do not readily generalize to deal with RDC measurements that come from different inner product spaces. Such complication gives rise to the open problem in \cite{donald2011algorithms}. 

We deal with it by introducing every monomial of the atom coordinates to our optimization problems instead of just using the elements of the Gram matrix. Furthermore, in our approach we view the protein backbone as an articulated structure composed of rigid planes and bodies that are chained together via hinges \cite{guntert1997torsion}, rather than just a loose set of points. The coordinates of the atoms can thus be determined by the orthogonal transformation of these rigid units. This has the advantage of lowering the
number of variables, and facilitating the incorporation of chirality constraint for the rigid units via re-parameterizing the problem in unit quaternion. We remark that 
(unlike existing optimization approaches that also model the protein as an articulated structure using torsion angle parameterization \cite{guntert1997torsion}, with RDC measurements alone) the cost and the constraints in our formulation are separable in the optimization variables (the unit quaternions), i.e. each term in the cost and constraints only depends on a single unit quaternion. Such structure of cost and constraints yields a less nonlinear optimization problem, which is essential in obtaining a convex relaxation to it.

%The constraints we described so far are in terms of the Cartesian coordinates of the atoms. As we will see in later sections, the atom coordinates can therefore be expressed in terms of rotations associated with the rigid units.  The determination of the rotations from RDC and NOE then provides the protein structure. 
		
Since the cost function (\ref{RDC equality}) is a 4-th order polynomial in the atomic coordinates (and in the rotations of the articulated structure as well), parameterizing the cost (\ref{RDC equality}) in unit quaternions gives rise to an 8-th order polynomial minimization problem. We introduce the SOS hierarchy \cite{lasserre2015introduction,blekherman2011semidefinite,lasserre2001global,parrilo2003semidefinite} to convexly relax the problem. The global optimum of the SOS relaxation is then used as our solution. One of the main benefit of such optimization technique is that when a solution is returned, whether or not the solution is the global optimum can be checked easily by examining the rank of certain matrices. Moreover, if we increase the complexity of the convex relaxation in the SOS hierarchy, the solution provably converges to the optimum of the original problem \cite{lasserre2001global}. Since its introduction, SOS relaxation has been applied successfully to solve many instances of non-convex optimization problems in sensor network localization \cite{nie2009sum}, super-resolution \cite{de2017exact}, tensor decomposition \cite{shah2015guaranteed}, and control theory \cite{henrion2005positive}. SOS optimization works by reformulating the polynomial optimization problem in terms of nonnegative polynomials. Since it is NP-hard to check whether a polynomial is nonnegative
%, in SOS relaxations sum-of-squares polynomials \cite{lasserre2015introduction} are searched for instead. 
the sum-of-squares polynominals \cite{lasserre2015introduction} are search in the SOS relaxations.
Empirically, when there is sufficient number of measurements, our proposed methods recover the optimal solution exactly and efficiently when there is no noise in the RDC, and stably when noise is added to the RDC, using low complexity SOS relaxation. 

The resulting algorithm RDC-SOS has running time of about an order of magnitude faster than existing toolboxes that use RDC for \emph{de novo} calculation of the protein backbone \cite{bryson2008redcraft,yershova2011rdc}. This is rather remarkable as the computational problem of determining the orientations has its domain on the product manifold of special orthogonal matrices, with a search space that is non-convex and exponential in size. Such fast and accurate determination of the initial structure could have potential applications in quick validation of backbone and NOE resonance assignment \cite{guntert2004cyana,zeng2009high} or refining Saupe tensor estimate through alternating minimization.

To include both RDC and NOE restraints to improve the solution quality, we propose a different method - RDC-NOE-SOS, at the expense of increasing the running time. 
%The NOE restraint has a very different flavor compare to RDC restraint in that it relates the rotations of two different rigid units, whereas RDC restricts the orientation of individual rigid unit in the Saupe tensor frame.
The nature of the NOE restraint is very different from an RDC constrain in that it indicates distances, translatable to rotation about hingers of the rigid units, whereas the RDC refers to the orientation of the individual rigid unit in the Saupe tensor frams.  
Therefore, when dealing with NOE restraints, only the relative rotations between the rigid units, i.e. the Gram matrix of the rotations, are optimized over. To incorporate these two types of measurement, our proposed method uses the information from RDC to regulate the spectrum of the rotation Gram matrix through a linear matrix inequality \cite{boyd2004convex}. The uses of rotation Gram matrix in RDC-NOE-SOS leads to a longer running time than RDC-SOS.

It is in general difficult to determine the backbone structure of an entire protein at once using an RDC-based algorithm, since along the chain of rigid units there are typically some sites having only a few or no RDC being measured. Therefore we divide up the protein backbone and run RDC-SOS or RDC-NOE-SOS on each of the fragments. As a separate contribution, we propose an additional SDP that jointly solves for the relative translations of all fragments using inter-fragment NOE in order to form the global structure of the protein. In \cite{yershova2011rdc}, a grid search is employed to find the translation that satisfies the NOE restraints between two fragments and the backbone is greedily and sequentially constructed based on the estimated pairwise translations. Our method, on the other hand, pieces all fragments at once rather than sequentially, and may therefore require fewer NOE measurements.

We tested the algorithms in calculating the structure of ubiquitin fragments from experimental RDC and NOE data deposited in the Protein Data Bank (PDB). We successfully computed the backbone structure for short fragments of ubiquitin (each consisting of 12 amino acids on average) up to 0.4\ \AA \ resolution, and the full backbone up to 0.86\ \AA\ resolution. This is competitive when comparing to the state of the art MFR method that gives structures with 0.56\ \AA\ and 0.87\ \AA\ RMSD for the fragments and full backbone of ubiquitin respectively. To further assess the quality of our structural calculation procedure, we introduce a classical statistical tool, the Cram\'er-Rao lower bound, which provides the minimum possible variance of the estimated atomic coordinates for a given noise model on the RDC and NOE. Both methods attain the CRB when aided by NOE restraints.

%DC added  	Fushman D, Ghose R, Cowburn D. The effect of finite sampling on the determination of orientational properties: A theoretical treatment with application to interatomic vectors in proteins. Journal of the American Chemical Society. 2000;122(43):10640-9  10.1021/ja001128j
%Even with multiple RDCs in different orientating media, the potential for non-unique solutions for the entire protein backbone remains \cite{fushman2000effect}.

\subsection{Broader contexts beyond structural biology}
In a broader context, our solution to the protein structuring problem presents a general strategy for determining the pose of an articulated structure, a common problem that arises in robotics and computer vision \cite{gavrila1999visual,andriluka2009pictorial}. The way we model the articulated structure from rotation matrices results in a cost function and constraints that are separable in the rotations, which in turn facilitates subsequent optimization. More generally, the SOS techniques used in our method could be applied to optimization problems involving low degree polynomials in terms of rotation matrices.

\subsection{Organization}
The rest of the paper is organized as follows. In Section \ref{section:Problem description}, we formulate the problem of backbone structure determination from RDC and NOE as a problem of finding the pose of an articulated structure. In Section \ref{section:NOE relaxation}, we describe a semidefinite program (SDP) that parallels the SDP proposed in \cite{so2007snl} for protein structuring from NOE in terms of rotation matrices. In Section \ref{section:RDC relaxation}, we apply the SOS relaxation to solve a general optimization problem involving polynomials of rotation matrices, which includes the structure determination problem from RDC. In Section \ref{section:NOE RDC relaxation}, we combine the two convex programs proposed in Section \ref{section:NOE relaxation} and Section \ref{section:RDC relaxation} to determine the pose of an articulated structure from both NOE and RDC.
In Section \ref{section:translation}, we propose an alternate SDP to piece together the fragments, when estimating the full protein structure directly is difficult. In Section \ref{section:numerical}, we present the numerical results with synthetic data and also for experimental data of ubiquitin (PDB ID: 1D3Z \cite{cornilescu19981d3z}).
%DC and will add other UB structure comparisonms?
In the appendix, we give a brief description of the RDC, the SOS relaxation, and we introduce the Cram\'er-Rao lower bound for the structure determination problem from RDC.

\subsection{Notation}
We use $\bm{I}_d$ to denote the identity matrix of size $d \times d$. We frequently use block matrices built from smaller matrices. For a block matrix $\bm{A}$, we use $\bm{A}_{ij}$ to denote its $(i,j)$-th block, $\bm{A}(p,q)$ to denote its $(p,q)$-th element, and $\bm{A}_i$ to denote the $i$-th column of $\bm{A}$. The size of the blocks will be made clear from the context. %We also employ the MATLAB \cite{guide1998matlab} notational convention, $A(p:q,r:s)$ to denote the matrix sampled from the $p$-th to $q$-th row and $r$-th to $s$-th column of $A$. $A(:,r:s)$ is used to denote the matrix sampled from $r$-th to $s$-th column of $A$.
We say that $\bm{A}$ is positive semidefinite (PSD) if $\bm{u}^T \bm{A} \bm{u} \geq 0$ for all $\bm{u}$, and use $\bm{A} \succeq \bm{B}$ to denote that $\bm{A} - \bm{B}$ is PSD \cite{boyd2004convex}, that is, . We use $\mathbb{O}(d)$ to denote the group of $d\times d$ orthogonal matrices. We use $\lVert \bm{x} \rVert_2$ to denote the Euclidean norm of $\bm{x} \in \mathbb{R}^n$ ($n$ should be clear from the context). We use $\mathrm{vec}(\bm{A})$ to denote the vectorization of a matrix $\bm{A}$, and $\mathrm{mat}(\bm{a})$ to denote the inverse procedure. In this paper we only use the $\mathrm{mat}(\cdot)$ operation to form a $3\times 3$ matrix from a column vector in $\mathbb{R}^9$. We denote the trace of a square matrix $\bm{A}$ by $\Tr(\bm{A})$. The Kronecker product between matrices $\bm{A}$ and $\bm{B}$ is denoted by $\bm{A} \otimes \bm{B}$. The all-ones vector and all-zeros vector are denoted by $\bm{1}$ and $\bm{0}$ respectively (the dimension should be obvious from the context). The $i$-th canonical basis vector is denoted as $\bm{e}_i$.

\section{Problem Formulation}
\label{section:Problem description}
In this section, we formulate the protein structuring problem as a non-convex optimization problem in terms of rotation matrices. The protein is composed of small rigid units whose structure is known, and we express the Cartesian coordinates of the atoms as well as RDC and NOE restraints in terms of the unknown rotation matrices associated with the rigid units. 

\subsection{Articulated structure and protein backbone}

\begin{figure}[h!]
  \centering
 \begin{subfigure}[b]{.8\linewidth}\centering\includegraphics[width=1\textwidth]{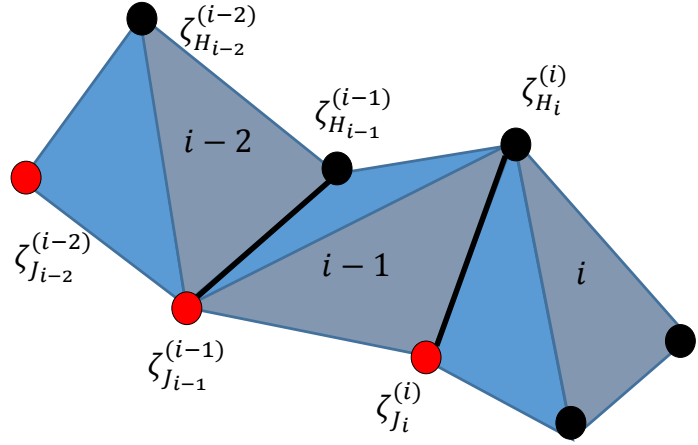}\caption{\centering}\end{subfigure}
 \begin{subfigure}[b]{.9\linewidth}\centering\includegraphics[width=1\textwidth]{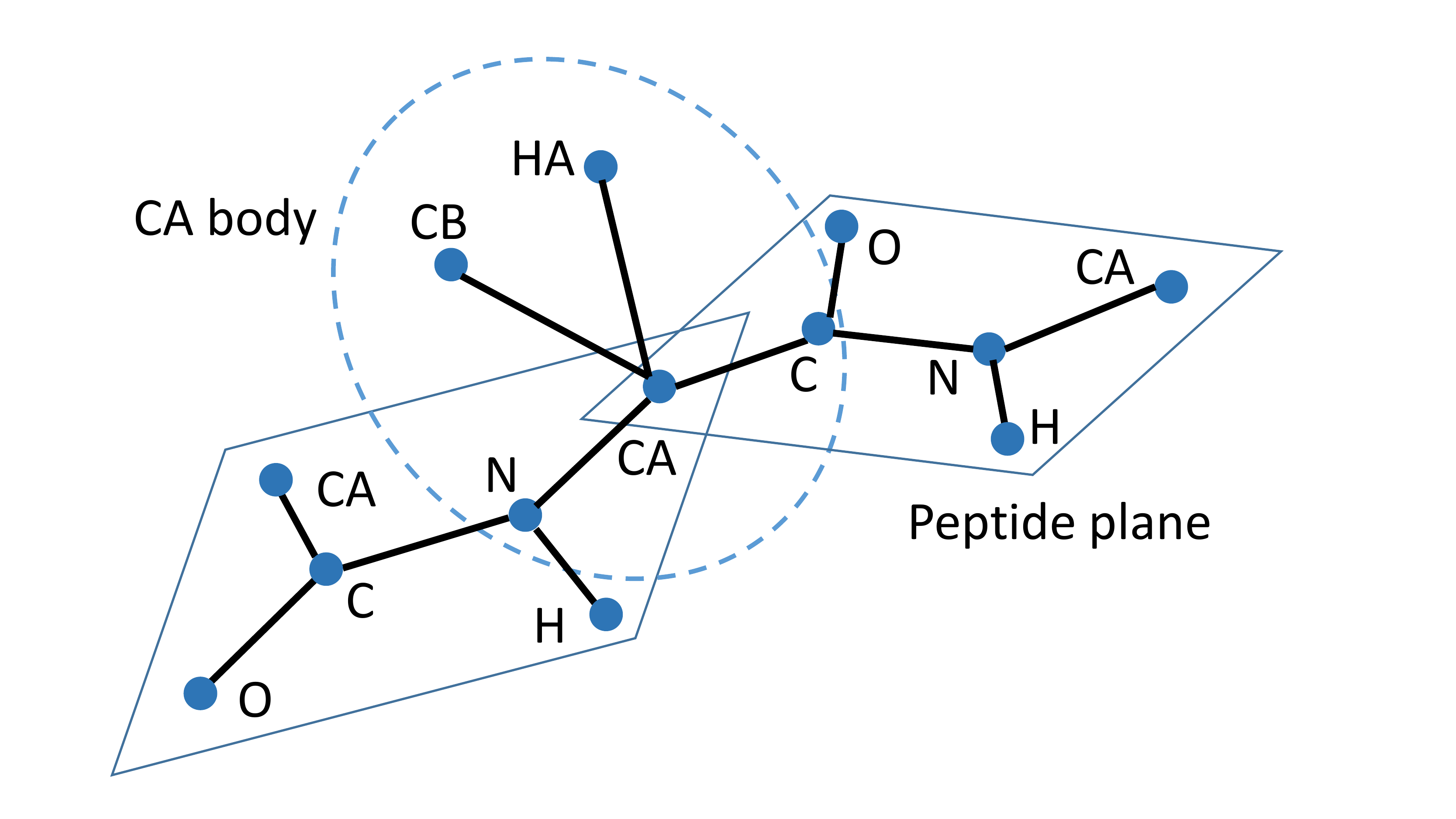}\caption{\centering}\end{subfigure}
    \caption{(a): Example of an articulated structure with joints with indices $J_i$'s (Red dots) and $H_i$'s. The hinges are represented by black bars in the figure. (b): Protein backbone consists of peptide planes and CA bodies. These rigid units are chained together at the bonds (N, CA) and (C,CA). }\label{figure:backbone}
\end{figure}

An articulated structure is a chain of rigid units where one unit is ``chained'' together with the next unit with non- overlapping joints (Figure \ref{figure:backbone}a). When there is a joint between two consecutive units, the relative translation is fixed but not the relative rotation. If there are two non-overlapping joints between two consecutive units, there is only one undetermined degree of freedom corresponding to a rotation around the axis defined by the two joints. This structure is also referred to as the \emph{body-hinge} framework \cite{whiteley2005counting} in rigidity theory. Let an articulated structure be composed of $K$ points residing in $M$ rigid units. For such a structure, we define a set of points $\{J_i\}_{i=1}^M$ as the joints between the units where $J_i \in \{1,\ldots,K\}$. The $i$-th unit is joined to the $(i-1)$-th unit at $J_i$.  Since the coordinates in each unit are known a-priori up to a rigid transformation, we then use $\bm{x}_k^{(i)}$ to denote the location of point $k$ in the local coordinate system of the $i$-th rigid unit. Notice that due to the rigid motion ambiguity, a Euclidean transform needs to be applied to each of the local coordinates $\bm{x}_k^{(i)}$ for each $i$ in order to form the global structure.

Let $\bm{\zeta}_k^{(i)}$ be the global coordinate of point $k$ in the $i$-th unit. For an articulated structure, it is possible to represent the global coordinates $\bm{\zeta}_k^{(i)}$ using the rotations $\bm{R}_i, i=1,\ldots,M$ associated with the $M$ rigid units. For $i=1$, we let
\begin{eqnarray}
\bm{\zeta}_k^{(1)} &=& \bm{R}_1 (\bm{x}^{(1)}_k-\bm{x}^{(1)}_{J_1})+\bm{t}
\end{eqnarray}
which amounts to orienting the first rigid unit with $\bm{R}_1$ and adding a translation so that $\bm{\zeta}^{(1)}_{J_1}$ are placed at $\bf{t}\in \mathbb{R}^3$. The coordinates for the $i=2$ rigid unit can be obtained as
\begin{eqnarray}
\bm{\zeta}_k^{(2)} &=& \bm{R}_2 (\bm{x}^{(2)}_k-\bm{x}^{(2)}_{J_2}) + \bm{\zeta}^{(1)}_{J_2}.
\end{eqnarray}
The above operations ensure that the $i=2$ rigid unit is jointed to the $i=1$ rigid unit at joint $J_2$, since $\bm{\zeta}^{(2)}_{J_2} = \bm{\zeta}^{(1)}_{J_2}$. The same reasoning implies that in general
\begin{eqnarray}
\label{recursion}
\bm{\zeta}_k^{(i)} &=& \bm{R}_i (\bm{x}_k^{(i)}-\bm{x}^{(i)}_{J_i}) + \bm{\zeta}^{(i-1)}_{J_i}.
\end{eqnarray}
Applying induction to (\ref{recursion}) gives
\begin{eqnarray}
\label{induction}
\bm{\zeta}_k^{(i)} &=&  \bm{R}_i (\bm{x}_k^{(i)}-\bm{x}^{(i)}_{J_i})  + \sum_{s=1}^{i-1} \bm{R}_s(\bm{x}^{(s)}_{J_{s+1}}-\bm{x}^{(s)}_{J_s}) + \bm{t}.
\end{eqnarray}
The coordinate of each atom is thus expressed as a linear combination of the rotations $\bm{R}_i$'s and a global translation $t$. As mentioned previously, when there are hinges in the articulated structure the rotations have fewer degrees of freedom. To incorporate the hinges, we define another set of joints $\{H_i\}_{i=1}^M$ where $\{H_i\}_{i=1}^M \cap \{J_i\}_{i=1}^M = \emptyset$. Let $\bm{v}_{kl}^{(i)}$ be the unit vector between the pair of points $(k,l)$ pointing from atom $l$ to atom $k$ in the frame of the $i$-th rigid unit. To ensure two consecutive rigid bodies stay chained together by a hinge, $\bm{R}_i$'s should satisfy the hinge constraints
\begin{eqnarray}
\label{hinge constraint}
\bm{R}_i \bm{v}^{(i)}_{H_i J_i} = \bm{R}_{i-1} \bm{v}^{(i-1)}_{H_iJ_i},\quad i=2,\ldots,M.
\end{eqnarray}

%Here we note that for the rigid planar units with affine rank of 2, it is not necessary to consider $\bm{R}_i$ but simply $R_i(:,1:2)$ that is on the Stiefel manifold $\mathbb{S}(3,2)$. This is due to the fact that we can assume the rigid planes lie in the $xy$-plane before re-orienting using $\bm{R}_i$. In this case, $\bm{x}_k^{(i)}(3) = 0$ for the local coordinate $\bm{x}_k^{(i)}$ and
%\begin{equation}
%\label{Stiefel}
%R_i \bm{x}_k^{(i)} = R_i(:,1:2) \bm{x}_k^{(i)}(1:2),
%\end{equation}
%rendering the third column of $\bm{R}_i$'s immaterial.

Using the above framework, we can reduce the problem of finding atomic coordinates of a protein backbone into a problem of finding the rotation matrices. This is because the protein backbone can be modeled as an articulated structure composed of peptide planes and CA-bodies. As depicted in Figure \ref{figure:backbone}b, a peptide plane is a 2D rigid plane consisting of atoms from two consecutive amino acids: $\text{CA}, \text{C}, \text{O}$ from one amino acid and $\text{H}, \text{N}, \text{CA}$ from the next amino acid. The CA-body is a 3D rigid body consisting of five atoms $\text{CA}, \text{N}, \text{C}, \text{HA}$ and $\text{CB}$ all coming from one amino acid. The bonds (N, CA), (C, CA) act like hinges between the rigid units.  
%DC add
This use of a rigid model requires no variation of the ${\omega}$ backbone torsion angle. 

%As mentioned earlier, the coordinates $\bm{x}_k^{(i)}$ and $y_k^{(i)}$ need to be re-oriented using rigid motions in order estimate the global backbone conformation. Let $R_i, P_i$ denote the orthogonal transformations that align the $i$-th CA-body and the $i$-th peptide plane. In the following subsections, we cast the backbone structural determination problem in terms of estimating these $\bm{R}_i$'s and $P_i$'s. In particular, instead of placing constraints concerning RDC and NOE on the coordinates directly as in (\ref{distance constraint}) and (\ref{rdc constraint}), we rewrite these constraints in terms of the orthogonal transformations $\bm{R}_i$.

\subsection{RDC data}
In the setting of calculating protein structure, the RDC measurements described in eq. \ref{rdc constraint} can be used to constrain the rotation of each rigid unit. Within each rigid unit, in principle all pairs of
% isotope-labeled atoms except those involving oxygen, O,   %% dc it's more complex than just not O
NMR active nuclei can give rise to an RDC, although in practice only a subset of these pairs have their RDC measured. Suppose $N$ Saupe tensors for the protein in $N$ different alignment media have been predetermined. In the $j$-th alignment media, the RDC measurements for the $i$-th rigid unit between the pair of atoms $(n,m)$, denoted $r_{nm}^{(j)}$, can be modeled in the following way:
\begin{eqnarray}
\label{RDC equality}
&r_{nm}^{(j)}& = {\bm{v}^{(i)}_{nm}}^T \bm{R}_i^T \bm{S}^{(j)} \bm{R}_i \bm{v}^{(i)}_{nm},\quad (n,m)\in {E_\mathrm{RDC}}_i,\cr
&\ &i=1,\ldots,M,\quad j=1,\ldots,N.
\end{eqnarray}
The set ${E_\mathrm{RDC}}_i$ is the set of edges that give rise to RDC in the $i$-th rigid unit, and $\bm{S}^{(j)}$ denotes the Saupe tensor in alignment media $j$. The orientation of the peptide planes and CA-bodies can be obtained by solving equation (\ref{RDC equality}) subject to the hinge constraint (\ref{hinge constraint}). Due to experimental errors in measuring the RDC, (\ref{RDC equality}) is only satisfied approximately, and orientations can be estimated by minimizing the following cost
\begin{multline}
\label{RDC minimization}
\sum_{i=1}^M \sum_{j=1}^N \sum_{(n,m)\in {E_\mathrm{RDC}}_i}\vert{\bm{v}^{(i)}_{nm}}^T \bm{R}_i^T \bm{S}^{(j)} \bm{R}_i \bm{v}^{(i)}_{nm} - r_{nm}^{(j)}\vert^2
\end{multline}
subject to (\ref{hinge constraint}).
In the cost function (\ref{RDC minimization}) each bond is counted once, including bonds that lie in both the peptide plane and the CA-body (e.g., bond $(\text{C} - \text{CA})$). The difficulty of minimizing the target function (\ref{RDC minimization}) lies in the non-convex nature of both the cost and domain.
Therefore, RDC measurements are typically used when refining an existing, high quality structure derived from solving the distance geometry problem from NOE or from homology modeling \cite{chen2012rdc}.

\subsection{NOE data}
\label{section:NOE data}

We now rewrite the distance constraints in (\ref{distance constraint}) in terms of the rotations. Instead of working with bounds on distances, we use bounds on squared distances, for reasons that will become apparent later in Section \ref{section:NOE relaxation}. Assuming $i>j$, from (\ref{induction}) we have
\begin{multline}
\label{NOE ortho}
\| \bm{\zeta}^{(i)}_m - \bm{\zeta}^{(j)}_n \|_2^2 =\|  \bm{R}_i (\bm{x}^{(i)}_m-\bm{x}^{(i)}_{J_i}) - \bm{R}_j (\bm{x}^{(j)}_n-\bm{x}^{(j)}_{J_j})\\
 + \sum_{s=j+1}^{i-1} \bm{R}_s (\bm{x}^{(s)}_{J_{s+1}}-\bm{x}^{(s)}_{J_s})\|_2^2.
\end{multline}
In this way, we write squared distances between two atoms as quadratic functions of $\bm{R}_i$'s. To satisfy the constraint (\ref{distance constraint}), we can minimize
\begin{multline}
\label{NOE unrelaxed}
\max( (d^\mathrm{low}_{mn})^2-\| \bm{\zeta}^{(i)}_m - \bm{\zeta}^{(j)}_n \|_2^2,0)^p + \\
\max(\| \bm{\zeta}^{(i)}_m - \bm{\zeta}^{(j)}_n \|_2^2 - (d^\mathrm{up}_{mn})^2,0)^p
\end{multline}
where the choice of the parameter $p$ depends on the specific noise model, and typical choices are $p=2$ (least squares) and $p=1$ (least unsquared deviations). In practice, the NOE measurements for the backbone
% atoms are more reliable and can also be treated as relatively hard constraints. %%DC
amide hydrigens are more reliable and yeild relatively hard constraints. 

When we have both RDC and NOE data, we simply combine (\ref{RDC minimization}) and (\ref{NOE unrelaxed})  into
\begin{multline}
\lambda \sum_{(m,n)\in E_\text{NOE}}\big[\max( (d^\mathrm{low}_{mn})^2-\| \bm{\zeta}^{(i)}_m - \bm{\zeta}^{(j)}_n \|_2^2,0)^p + \\
\max(\| \bm{\zeta}^{(i)}_m - \bm{\zeta}^{(j)}_n \|_2^2 - (d^\mathrm{up}_{mn})^2,0)^p \big]+\\
\sum_{i=1}^M \sum_{j=1}^N \sum_{(n,m)\in {E_\mathrm{RDC}}_i}\vert{\bm{v}^{(i)}_{nm}}^T \bm{R}_i^T \bm{S}^{(j)} \bm{R}_i \bm{v}^{(i)}_{nm} - r_{nm}^{(j)}\vert^2
\end{multline}
where $E_\text{NOE}$ denotes the set of atom pairs that have NOE measured. The choice of $\lambda$ is typically around $10^{-9}$.

\section{Convex relaxation with only NOE constraints}\label{section:NOE relaxation}
In this section, we describe a convex relaxation to solve the non-convex protein structuring problem from NOE. We purposefully choose not to present the convex relaxation for RDC first because the concepts involved there are more complicated, and presenting the simpler case with only NOE data can help readers to develop intuitions. 

In order to deal with NOE restraints, we first write the problem in terms of the Gram matrix of rotations, i.e. the relative rotation $\R_i^T \R_j$ for every pair of $(i,j),\ i,j=1,\ldots,M$. This is made possible since the Euclidean distances between different atoms only depend on the inner products of the atom coordinates. From (\ref{NOE ortho}), we get
\begin{multline}
\| \bm{\zeta}^{(i)}_m - \bm{\zeta}^{(j)}_n \|_2^2 = \Tr\bigg([\bm{R}_i,\bm{R}_{i+1},\ldots,\bm{R}_j]^T [\bm{R}_i,\bm{R}_{i+1},\ldots,\bm{R}_j]\cr
\begin{bmatrix}
\x^{(j)}_n - \x^{(j)}_{J_j}\\ \x^{(j+1)}_{J_{j+1}} - \x^{(j+1)}_{J_j}\\ \vdots \\ \x^{(i)}_m - \x^{(i)}_{J_i}
\end{bmatrix} \begin{bmatrix}
\x^{(j)}_n - \x^{(j)}_{J_j}\\ \x^{(j+1)}_{J_{j+1}} - \x^{(j+1)}_{J_j}\\ \vdots \\ \x^{(i)}_m - \x^{(i)}_{J_i}
\end{bmatrix}^T \bigg).
\end{multline}
Introducing a new matrix variable
\begin{equation}
\label{G form}
\G = [\bm{R}_1,\ldots,\bm{R}_M]^T [\bm{R}_1,\ldots,\bm{R}_M] \in \mathbb{R}^{3M\times 3M},
\end{equation}
the cost (\ref{NOE unrelaxed}) can be written as
\begin{multline}
\label{NOE in P1}
f_\text{NOE}(\bm{G}) = \sum_{(m,n)\in E_\text{NOE}}\max( (d^\mathrm{low}_{mn})^2-\Tr(\bm{A}_{mn} \bm{G}),0)^p +\cr
 \max( \Tr(\bm{A}_{mn} \bm{G}) - (d^\mathrm{up}_{mn})^2,0)^p
\end{multline}
using some coefficient matrices $\bm{A}_{mn}$'s. In terms of this new variable $\G$, we formulate the minimization problem
\begin{eqnarray}
\label{NOE with G}
&\ &\min_{\bm{G}}  f_\text{NOE}(\bm{G}) \\
&\mathrm{s.t.} & \bm{G}_{ii} = \bm{I}_3,\cr
&\ & \bm{G}\succeq 0,\cr
&\ & \text{rank}(\bm{G}) = 3,\cr
&\ & \bm{v}_{J_i H_i}^{(i-1)} = \bm{G}_{(i-1)i} \bm{v}_{J_i H_i}^{(i)},\ i\in[2,M],\nonumber
\end{eqnarray}
where $\G_{ii}$ denotes the $3\times 3$ blocks on the diagonal of the matrix $\G$. The first three constraints are equivalent to (\ref{G form}), in particular, the PSD-ness and rank-3 constraints of $\G$ ensures the existence of a factorization in (\ref{G form}) and $\G_{ii} = \bm{I}_3$ is equivalent to the orthogonality of $\R_i$'s. The last constraint comes from (\ref{hinge constraint}), by changing $\R_{i-1}^T \R_{i}$ to $\G_{(i-1)i}$. Problem (\ref{NOE with G}) is (almost) equivalent to the problem of finding the chain of rotations from NOE data, except it does not consider the chirality constraint $\det(\R_i)>0$.

Observe that the cost in (\ref{NOE with G}) is convex in the variable $\G$, whereas the domain is non-convex due to the rank-3 constraint. We therefore drop the rank-3 constraint in order to derive a convex relaxation that is similar to the Max-Cut \cite{goemans1995improved} and rotation synchronization \cite{singer2011angular,cucuringu2012eigenvector} SDP relaxations:
\begin{eqnarray}
\label{NOE with G relaxed}
&\ \ &\min_{\bm{G}}  f_\text{NOE}(\bm{G}) \\
&\mathrm{s.t.} & \bm{G}_{ii} = \bm{I}_3,\cr
&\ & \bm{G}\succeq 0,\cr
&\ & \bm{v}_{J_i H_i}^{(i-1)} = \bm{G}_{(i-1)i} \bm{v}_{J_i H_i}^{(i)},\ i\in[2,M].\nonumber
\end{eqnarray}
This SDP can be easily solved using, e.g., one of the solvers implemented in CVX \cite{cvx}, a library of conic-programs solver available in Matlab and Python. There is a natural interpretation of the rank relaxed problem (\ref{NOE with G relaxed}) that is similar to the SDP proposed in \cite{so2007snl}, in which the orthogonal transformation associated with each rigid unit is in a high dimensional ambient space $\mathbb{R}^{3M}$ (instead of $\mathbb{R}^3$). To see this, since $\G\succeq 0$ in (\ref{NOE with G relaxed}), it admits a Cholesky factorization
\begin{equation}
\G = [\bm{P}_1,\cdots,\bm{P}_M]^T [\bm{P}_1,\cdots,\bm{P}_M]
\end{equation}
where $\bm{P}_i\in \mathbb{R}^{3M\times 3}$. Furthermore, since $\G_{ii}=\bm{I}_3$, $\bm{P}_i^T \bm{P}_i = \bm{I}_3$. Therefore
\begin{equation}
\bm{\zeta}^{(i)}_k = \bm{P}_i \x^{(i)}_k
\end{equation}
is a vector in $\mathbb{R}^{3M}$, obtained from rigidly transforming $\x^{(i)}_k\in \mathbb{R}^3$ into $\bm{\zeta^{(i)}}_k\in \mathbb{R}^{3M}$. When applying (\ref{recursion}) with $\bm{P}_i$ instead of $\R_i$, we have a framework in $\mathbb{R}^{3M}$ instead of $\mathbb{R}^3$, and the NOE constraint (\ref{NOE unrelaxed}) is now a distance constraint placed on a body-hinge framework in $\mathbb{R}^{3M}$.

While the global optimum of (\ref{NOE with G relaxed}) can be efficiently obtained in polynomial time through standard convex optimization methods, its solution will only resemble the solution of (\ref{NOE with G}) closely if there is a sufficient number of NOE constraints restricting the body-hinge framework to have a unique configuration in a low dimensional subspace close to $\mathbb{R}^3$ \cite{so2007theory}. Without sufficient distance measurements, the quality of the embedding can deteriorate quickly. In the next section, we show how one can use RDC measurements to further improve the quality of the solution using RDC.

\section{Convex relaxation with only RDC constraints}\label{section:RDC relaxation}
To determine the backbone structure from RDC, the cost (\ref{RDC minimization}) needs to be minimized. Unlike the case of NOE (\ref{NOE unrelaxed}), each term in (\ref{RDC minimization}) is a 4-th order polynomial function involving a single rotation. In Section  \ref{section:single rotation relaxation}, we first examine the case of optimization over a single rotation. We parameterize the rotations using unit quaternions and apply the SOS hierarchy for polynomial optimization problem over the set of unit quaternions. In Section \ref{section:size reduction}, we further discuss a few techniques to reduce the size of the proposed convex relaxations. Finally in Section \ref{section:multiple rotations}, based on the SOS formulation for optimization over a single rotation,  we propose a convex relaxation to jointly optimize multiple rotation matrices that are coupled through the hinge constraints (\ref{hinge constraint}). This leads to the algorithm - RDC-SOS, for protein structuring from RDC data.

\subsection{Optimization over a single rotation}\label{section:single rotation relaxation}
In order to handle RDC, as mentioned earlier we need to be able to optimize a fourth order polynomial in terms of rotations. We first derive a convex relaxation method for solving the non-convex optimization problem of the form
\begin{equation}
\label{general SO3 optimization}
\min_{\bm{R}\in\mathbb{SO}(3)} f(\bm{R}).
\end{equation}
where $f$ is a polynomial function. By the Euler-Rodrigues formula, a rotation matrix can be derived from a unit quaternion $\q\in \mathbb{R}^4$ via
\begingroup\makeatletter\def\f@size{9}\check@mathfonts
\begin{multline}
\label{quat2rot}
\R = \phi(\q \q^T):=\\
\left[\begin{smallmatrix} 1-2 \bm{q}(3)^2 - 2 \bm{q}(4)^2 & 2(\bm{q}(2)\bm{q}(3) - \bm{q}(4)\bm{q}(1)) & 2(\bm{q}(2)\bm{q}(4) + \bm{q}(3)\bm{q}(1))\\
2(\bm{q}(2)\bm{q}(3) + \bm{q}(4)\bm{q}(1)) & 1-2 \bm{q}(2)^2 - 2 \bm{q}(4)^2  & 2 (\bm{q}(3)\bm{q}(4)-\bm{q}(2)\bm{q}(1))\\
 2(\bm{q}(2)\bm{q}(4) - \bm{q}(3)\bm{q}(1)) & 2 (\bm{q}(3)\bm{q}(4)+\bm{q}(2)\bm{q}(1)) & 1 - 2\bm{q}(2)^2-2\bm{q}(3)^2\end{smallmatrix}\right]
\end{multline}
\endgroup
where $\q^T \q = 1$. Therefore, we can equivalently consider solving an optimization problem of the form:
\begin{equation}
\label{general quat optimization}
\min_{\bm{q}: \bm{q}^T \bm{q}=1} f(\phi(\q \q^T)).
\end{equation}

The choice of such parametrization using unit quaternion is motivated by the following fact. Let us consider the easier problem of optimizing a linear function over the set of rotation matrices. For any matrix $\bm{C}\in\mathbb{R}^{3\times 3}$,
\begin{eqnarray}
\min_{\bm{R}\in\mathbb{SO}(3)} \Tr(\bm{C} \R) &=& \min_{\bm{q}\in\mathbb{R}^4:\ \bm{q}^T \bm{q}=1} \Tr(\bm{C} \phi(\q \q^T)) \cr
&=& \min_{\bm{Q}\in\mathbb{R}^4 :\ \Tr(\bm{Q})=1, \bm{Q}\succeq 0  } \Tr(\bm{C} \phi(\bm{Q})).
\end{eqnarray}
Denoting the convex hull of a set $\mathcal{S}$ as $\text{conv}(\mathcal{S})$, the last equality comes from the fact that
\begin{multline}
\text{conv}(\{\q \q^T\ \vert\ \q^T \q = 1,\ \q\in\mathbb{R}^4\}) \cr
= \{\bm{Q}\in\mathbb{R}^{4\times 4}\ \vert\ \Tr(\bm{Q}) = 1, \bm{Q}\succeq 0\}.
\end{multline}
By relaxing the domain of the set of matrices $\q \q^T$ to its convex hull, optimizing a linear function over the special orthogonal group can be done exactly. Motivated by this observation, we may hope a suitable convex relaxation on the set of monomials of the unit quaternion to yield a tight convex relaxation to problem (\ref{general SO3 optimization}) where $f$ is a higher-degree polynomial of the rotation matrices. We therefore consider solving (\ref{general quat optimization}) in place of (\ref{general SO3 optimization}).

In order to obtain a convex relaxation to (\ref{general quat optimization}), a general strategy-the SOS relaxation, can be used to derive hierarchies of convex relaxations with increasing complexity to solve problem (\ref{general SO3 optimization}). We give an intuitive presentation of the procedure here and refer interested reader to the appendix and excellent texts such as \cite{lasserre2015introduction,blekherman2011semidefinite} for a more in depth exposition. In the rest of the paper, for any variable $\bm{x}\in \mathbb{R}^n$ and $\bm{\alpha}\in \mathbb{N}^n$, $\bm{x}^{\bm{\alpha}}$ is defined as
\begin{equation}
\bm{x}^{\bm{\alpha}}:= \x(1)^{\bm{\alpha}(1)} \x(2)^{\bm{\alpha}(2)} \cdots \bm{x}(n)^{\bm{\alpha}(n)}
\end{equation}
The notations $[\bm{x}]_{\leq d}$ and $[\bm{x}]_d$ are used to denote vectors containing the monomials of $\bm{x}$ up to and with degree $d$ respectively. We remind the reader that for a $n$-dimensional variable, there are $\binom{n+d}{d}$ distinct monomials up to degree $d$, and $\binom{n+d-1}{d}$ degree $d$ monomials. We often use vector $\bm{x}$ with the size of $[\q]_{\leq d}$ ($[\q]_{d}$) or matrix $\bm{X}$ with the size of $[\q]_{\leq d}[\q]_{\leq d}^T$ ($[\q]_{d}[\q]_{d}^T$). In this case, we use $\bm{x}_{\bm{\alpha}}$ or $\bm{X}_{\bm{\alpha} \bm{\beta}}$ to denote the entries associated with $\q^{\bm{\alpha}}$ or $\q^{\bm{\alpha}} \q^{\bm{\beta}}$ respectively. For a vector $\bm{\alpha}\in \mathbb{N}^n$ of natural numbers, we define $\vert \bm{\alpha} \vert := \sum_{i=1}^n \bm{\alpha}(i)$. Finally, we say a polynomial $g(\bm{x})$ is t-SOS if $g(\bm{x}) = \sum_i h_i(\bm{x})^2$ where $h_i(\bm x)$'s are some polynomials with the highest degree being $t$.

Now, introducing matrix variables
\begin{equation}
\mathcal{M}_{\leq 2d} := [\q]_{\leq d} [\q]_{\leq d}^T,
\end{equation}
and
\begin{equation}
\mathcal{M}_{2d} := [\q]_{d} [\q]_{d}^T,
\end{equation}
and assuming $f$ is a polynomial of rotation $\bm{R}$ with degree at most $t$, problem (\ref{general quat optimization}) can be written equivalently as
\begin{eqnarray}
\label{general quat optimization 2}
&\ & \min_{\mathcal{M}_{\leq 2t}, [\q]_{\leq t}} \sum_{\vert\bm{\alpha}\vert\leq 2t} f_{\bm{\alpha}, \text{even}} \q^{\bm{\alpha}}\\
&\text{s.t.}&\ \mathcal{M}_{\leq 2t} = [\q]_{\leq t} [\q]_{\leq t}^T,\cr
&\ &{\mathcal{M}_{\leq 2t}}_{\bm{\alpha}\bm{\beta}}(\q^T\q -1) = 0,\ \text{if}\ \vert\bm{\alpha}+\bm{\beta}\vert \leq 2t-2.\label{redundant quat eq 0}
\end{eqnarray}
Here
\begin{equation}
f(\phi(\q\q^T)) := \sum_{\bm{\alpha}} f_{\bm{\alpha}, \text{even}} \q^{\bm{\alpha}}
\end{equation}
where the polynomial only involves even degree monomials. Notice that the coefficients $f_{\bm{\alpha}, \text{even}}$ that appears in (\ref{general quat optimization 2}) corresponds to $f(\phi(\q \q^T))$ in (\ref{general quat optimization}) rather than $f(\bm{R})$ in (\ref{general SO3 optimization}). We have inserted a set of equality constraints (\ref{redundant quat eq 0}) that seems redundant, as they are all implied by $\q^T \q - 1 = 0$. However, these are crucial if we want to obtain a convex relaxation of the non-convex problem (\ref{general quat optimization 2}). Observe that these equalities are linear in the variable $\mathcal{M}_{\leq 2t}$. At this point, the obstacle of having a convex problem is due to the nonlinear equality constraint
\begin{eqnarray}
\mathcal{M}_{\leq 2t} = [\q]_{\leq t} [\q]_{\leq t}^T,
\end{eqnarray}
which is equivalent to
\begin{gather}
\mathcal{M}_{\leq 2t}\succeq 0,\ {\mathcal{M}_{\leq 2t}}_{\bm{\alpha} \bm{\beta}} = \bm{y}_{\bm{\alpha}+\bm{\beta}},\ \bm{y}_{[0,0,0,0]} = 1,\\
\sum_{i=1}^4 \bm{y}_{\bm{\alpha}+ 2\bm{e}_i} - \bm{y}_{\bm{\alpha}} =0\ \text{if}\ \vert\bm{\alpha}\vert\leq 2t-2,\ \text{rank}(\mathcal{M}_{\leq 2t})=1,
\end{gather}
for some $\bm{y}\in \mathbb{R}^{\binom{n+2t}{2t}}$. If the non-convex rank constraint is removed, the convex problem
\begingroup\makeatletter\def\f@size{8}\check@mathfonts
\begin{eqnarray}
\label{general quat optimization 3}
&\ &\min_{\mathcal{M}_{\leq 2t},\bm{y}} \sum_{\vert \bm{\alpha}\vert \leq 2t} f_{\bm{\alpha}, \text{even}} \bm{y}_{\bm{\alpha}} \\
&\quad\text{s.t.} & \mathcal{M}_{\leq 2t}\succeq 0,\label{PSD moment}\\
&\ & \bm{y}_{[0,0,0,0]} = 1,\label{normalized moment}\\
&\ & {\mathcal{M}_{\leq 2t}}_{\bm{\alpha} \bm{\beta}} = \bm{y}_{\bm{\alpha}+\bm{\beta}},\label{same moment}\\\
&\ & \sum_{i=1}^4 \bm{y}_{\bm{\alpha}+ 2\bm{e}_i} - \bm{y}_{\bm{\alpha}} =0,\ \text{for all}\ \bm{\alpha}\in\mathbb{N}^4,\ \vert\bm{\alpha}\vert\leq 2t-2. \label{redundant quat eq}
\end{eqnarray}
\endgroup
is obtained. Here $\bm{e}_i$'s are canonical basis vectors in $\mathbb{R}^4$. The purpose of having the redundant constraints (\ref{redundant quat eq 0}) is now clear. When the rank-1 constraint of $\mathcal{M}_{\leq 2t}$ is removed, it is easy to verify that the entries of $\mathcal{M}_{\leq 2t}$ need not be high order monomials of $\q$. More precisely, $\bm{y}_{\bm{\alpha}+\bm{\beta}}\neq \bm{y}_{\bm{\alpha}}\bm{y}_{\bm{\beta}}$ generally, which means $\sum_{i=1}^4 \bm{y}_{2\bm{e}_i} - 1 =0$ does not imply $\sum_{i=1}^4 \bm{y}_{\bm{\alpha}+ 2\bm{e}_i} - \bm{y}_{\bm{\alpha}} =0$. Although (\ref{general quat optimization 3}) is a surrogate convex problem of (\ref{general quat optimization 2}), in \cite{lasserre2001global} it is shown that as $t$ tends to infinity, the optimum of problem (\ref{general quat optimization 3}) converges to problem (\ref{general quat optimization 2}). We observe in our numerical study that in many instances, problem (\ref{general quat optimization 3}) converges to (\ref{general quat optimization 2}) already for small and finite values $t$, as in many other applications of SOS relaxation \cite{nie2014optimality}.
%\label{key}

%since problem (\ref{general quat optimization 3}) is the dual problem to certain polynomial optimization problem. More precisely, if we want to solve
%\begin{equation}
%\min_{\bm{x}} f(\bm{x})\quad\text{s.t.}\ g(\bm{x}) = 0
%\end{equation}
%where $f(\bm{x})$ is a polynomial function, we can instead solve
%\begin{equation}
%\max_{\gamma} \gamma \quad\text{s.t.}\ f(\bm{x})-\gamma\geq 0\ \text{on}\ \{\bm{x}\vert g(\bm{x})= 0\},
%\end{equation}
%or
%\begin{equation}
%\label{nonnegative}
%\max_{\gamma,h(\bm{x})} \gamma \quad\text{s.t.}\ f(\bm{x})-\gamma - h(\bm{x}) g(\bm{x}) \geq 0.
%\end{equation}
%Here the optimization variables are constant $\gamma$ and the coefficients of the polynomial $h(\bm{x})$. In general, deciding the nonnegativity of a polynomial is an NP-hard problem, hence (\ref{nonnegative}) is generally intractable. However, since deciding whether a polynomial is a SOS is computationally tractable, we can solve
%\begin{equation}
%\max_{\gamma,h(\bm{x})} \gamma \quad\text{s.t.}\ f(\bm{x})-\gamma - h(\bm{x}) g(\bm{x})\ \text{is SOS}
%\end{equation}
%or equivalently,
%\begin{equation}
%\label{SOS}
%\max_{\gamma,h(\bm{x}),\bm{Q}\succeq 0} \gamma \quad\text{s.t.}\ f(\bm{x})-\gamma - h(\bm{x}) g(\bm{x}) = [\bm{x}]_t \bm{Q} [\bm{x}]_t.
%\end{equation}
%This is a standard semidefinite program. The dual problem to (\ref{SOS}) is indeed

\subsubsection{Variable size reduction}
\label{section:size reduction}
In problem (\ref{general quat optimization 3}), the matrix $\mathcal{M}_{\leq 2t}$ is of size $\binom{4+t}{t}\times \binom{4+t}{t}$. Here we exploit the special structure of the problem to reduce the size of the variable to $\binom{3+t}{t}\times \binom{3+t}{t}$, which is the size of the highest degree block, i.e. $\mathcal{M}_{2t}$ in the matrix $\mathcal{M}_{\leq 2t}$. For example, for the typical value $t=4$ the matrix size is reduced from $70\times 70$ to $35\times 35$.

The first observation is that since the set of unit quaternions forms a double cover of $\mathbb{SO}(3)$, i.e. both $\q$ and $-\q$ define the same rotation matrix, it is not necessary to consider odd degree monomials in $\mathcal{M}_{\leq 2t}$. More precisely, we arrange $\mathcal{M}_{\leq 2t}$ such that
\begin{equation}
\mathcal{M}_{\leq 2t} = \begin{bmatrix} \mathcal{M}_{\leq 2t,\text{even}} & \mathcal{M}_{\leq 2t,\text{odd}} \\ \mathcal{M}_{\leq 2t,\text{odd}}^T &  \mathcal{M}_{\leq 2(t-1),\text{even}}\end{bmatrix}
\end{equation}
where
\begin{equation}
\mathcal{M}_{\leq 2t, \text{even}} = \begin{bmatrix} [\q]_{0} \\ [\q]_2 \\ \vdots \\ [\q]_t \end{bmatrix} \begin{bmatrix} [\q]_{0}^T & [\q]_2^T & \cdots & [\q]_t^T \end{bmatrix},
\end{equation}
\begin{equation}
\mathcal{M}_{\leq 2t, \text{odd}} = \begin{bmatrix} [\q]_{0} \\ [\q]_2 \\ \vdots \\ [\q]_t \end{bmatrix} \begin{bmatrix} [\q]_{1}^T & [\q]_3^T & \cdots & [\q]_{t-1}^T \end{bmatrix},
\end{equation}
and
\begin{equation}
\mathcal{M}_{\leq 2(t-1), \text{even}} = \begin{bmatrix} [\q]_{1} \\ [\q]_3 \\ \vdots \\ [\q]_{t-1} \end{bmatrix} \begin{bmatrix} [\q]_{1}^T & [\q]_3^T & \cdots & [\q]_{t-1}^T \end{bmatrix}.
\end{equation}
After the convex relaxation in (\ref{general quat optimization 3}), we may assume $\mathcal{M}_{\leq 2t, \text{odd}}=0$. Due to the quadratic dependence of a rotation matrix on the unit quaternion, the cost of the optimization problem (\ref{general quat optimization 3}) only involves $\mathcal{M}_{\leq 2t, \text{even}}$ and $\mathcal{M}_{\leq 2(t-1), \text{even}}$, hence both matrices
\begin{equation}
\label{two solutions}
\begin{bmatrix} \mathcal{M}_{\leq 2t,\text{even}} &  \mathcal{M}_{\leq 2t,\text{odd}} \\  \mathcal{M}_{\leq 2t,\text{odd}}^T &  \mathcal{M}_{\leq 2(t-1),\text{even}}\end{bmatrix}, \ \begin{bmatrix} \mathcal{M}_{\leq 2t,\text{even}} &  -\mathcal{M}_{\leq 2t,\text{odd}} \\  -\mathcal{M}_{\leq 2t,\text{odd}}^T &  \mathcal{M}_{\leq 2(t-1),\text{even}}\end{bmatrix}
\end{equation}
give the same cost. Therefore without lost of generality, we may assume the solution we seek is the average of the matrices in (\ref{two solutions})
\begin{equation}
\mathcal{M}_{\leq 2t} = \begin{bmatrix} \mathcal{M}_{\leq 2t,\text{even}} & \bm{0} \\ \bm{0} &  \mathcal{M}_{\leq 2(t-1),\text{even}}\end{bmatrix}\succeq 0.
\end{equation}
Now, notice that the entries of $\mathcal{M}_{\leq 2(t-1), \text{even}}$ are a subset of the entries of $\mathcal{M}_{\leq 2t, \text{even}}$, it seems plausible that the constraint $\mathcal{M}_{\leq 2(t-1), \text{even}}\succeq 0$ in (\ref{two solutions}) can be dropped altogether and problem (\ref{general quat optimization 3}) can be written solely in terms of $\mathcal{M}_{\leq 2t,\text{even}}$. Indeed, the constraint $\mathcal{M}_{\leq 2(t-1), \text{even}}\succeq 0$ is already implied by $\mathcal{M}_{\leq 2t,\text{even}}\succeq 0$, since for odd $\bm{\alpha},\bm{\beta}$
\begin{equation}
{\mathcal{M}_{\leq 2(t-1), \text{even}}}_{\bm{\alpha}\bm{\beta}} = \sum_{k=1}^4 \bm{y}_{\bm{\alpha} +\bm{e}_k +\bm{\beta}+\bm{e}_k} =  \sum_{k=1}^4 {\mathcal{M}_{\leq 2t, \text{even}}}_{(\bm{\alpha}+\bm{e}_k) (\bm{\beta}+\bm{e}_k)}
\end{equation}
where the first equality follows from (\ref{redundant quat eq}) (in other words, ${\mathcal{M}_{\leq 2(t-1), \text{even}}} = \bm{A} {\mathcal{M}_{\leq 2t, \text{even}}} \bm{A}^T$ for some matrix $\bm{A}$). Let
\begin{equation}
p(d) := \sum_{i=0, i\ \text{even}}^d \binom{3+i}{i},
\end{equation}
where we note that the size of $\mathcal{M}_{\leq 2t, \text{even}}$ is $p(t)$.

The second size reduction comes from the equality constraints in line (\ref{redundant quat eq}). Essentially (\ref{redundant quat eq}) implies
\begin{multline}
\sum_{i=1}^4 {\mathcal{M}_{\leq 2t, \text{even}}}_{\bm{\alpha}+\bm{\beta}+2\bm{e}_i} -   {\mathcal{M}_{\leq 2t, \text{even}}}_{\bm{\alpha}+\bm{\beta}} \cr= 0,\ \text{if}\  \vert\bm{\alpha}+\bm{\beta}\vert \leq 2t-2,
\end{multline}
or equivalently,
\begingroup\makeatletter\def\f@size{8}\check@mathfonts
\begin{eqnarray}
&&\big(\bm{e}_{[2,0,0,0]+\bm{\alpha}}^T +\bm{e}_{[0,2,0,0]+\bm{\alpha}}^T+\bm{e}_{[0,0,2,0]+\bm{\alpha}}^T+\bm{e}_{[0,0,0,2]+\bm{\alpha}}^T 
- \bm{e}_{[0,0,0,0]+\bm{\alpha}}^T\big)\mathcal{M}_{\leq 2t, \text{even}}\cr 
&&= 0 \quad \forall\bm{\alpha}\ \text{such that}\ \vert \bm{\alpha} \vert \leq 2t-2.\label{facial reduction null}
\end{eqnarray}
\endgroup
where $\bm{e}_{\bm{\alpha}}$ for $\bm{\alpha}\in\mathbb{N}^4$ are canonical basis vectors with the size of $[\bm{q}]_{\leq t}$. Notice that there are $p(t-1)$ equality constraints on $\mathcal{M}_{\leq 2t}\leq 2t$ in (\ref{facial reduction null}). Let $\bm{U}_{\leq 2t}\in \mathbb{R}^{p(t)\times (p(t)-p(t-1))}$ be a matrix whose columns form a basis to $\text{Range}(\mathcal{M}_{\leq 2t, \text{even}})$,
\begin{equation}
\mathcal{M}_{\leq 2t, \text{even}} = \bm{U}_{\leq 2t} \tilde{\mathcal{M}}_{\leq 2t} \bm{U}_{\leq 2t}^T
\end{equation}
where $\tilde{\mathcal{M}}_{\leq 2t} $ has size $(p(t)-p(t-1))\times(p(t)-p(t-1)) = \binom{3+t}{t}\times\binom{3+t}{t}$ and is PSD. Hereafter, whenever we optimize over the variable $\mathcal{M}_{\leq 2t}$, we can remove all the equality constraints in (\ref{redundant quat eq}) and simply work with $\tilde{\mathcal{M}}_{\leq 2t}$. We note that this procedure of removing the nullspace of a PSD variable is called facial reduction \cite{krislock2010semidefinite} in the convex optimization community. Semidefinite facial reduction is not only important for reducing the size of the variable, but also necessary to ensure the numerical stability of semidefinite programs.

\subsection{Optimization over multiple rotations - RDC-SOS}
\label{section:multiple rotations}
When minimizing the RDC cost in (\ref{RDC minimization}) over multiple rotations, the hinge constraints (\ref{hinge constraint}) have to be included. In terms of unit quaternions, constraints (\ref{hinge constraint}) are of the form
\begin{equation}
\bm{l}_{1,i}(\phi(\q_i\q_i^T)) = \bm{l}_{2,i}(\phi(\q_{i-1}\q_{i-1}^T)),\ i=2,\ldots,M.
\end{equation}
where the linear functions $\bm{l}_{1,i},\bm{l}_{2,i}: \mathbb{R}^{3\times3} \rightarrow \mathbb{R}^{3}$. Again, we have to include other valid equalities
\begin{equation}
\label{hinge constraint 2}
\bm{l}_{1,i}(\phi(\q_i\q_i^T))^{\bm{\gamma}} = \bm{l}_{2,i}(\phi(\q_{i-1}\q_{i-1}^T))^{\bm{\gamma}},\ i=2,\ldots,M,
\end{equation}
as in Section \ref{section:single rotation relaxation} since we are going to apply SOS relaxation to the matrices $\mathcal{M}_{\leq 2t,i} = [\q]_{\leq t,i} [\q]_{\leq t,i}^T$.
Let
\begin{gather}
\bm{l}_{1,i}(\phi(\q_i\q_i^T))^{\bm{\gamma}} := \sum_{\bm \alpha} {\bm{l}_{1,i}^{\bm{\gamma}}}_{\bm \alpha} \q_i^{\bm{\alpha}},\cr
\bm{l}_{2,i}(\phi(\q_i\q_i^T))^{\bm{\gamma}} := \sum_{\bm \alpha} {\bm{l}_{2,i}^{\bm{\gamma}}}_{\bm \alpha} \q_i^{\bm{\alpha}},\cr
i=2,\ldots,M,
\end{gather}
Also, we define for each term of the cost (\ref{RDC minimization}) its polynomial expansion in terms of its corresponding quaternion
\begin{multline}
\sum_{\bm{\alpha}} {f_{\text{RDC},\bm{\alpha}}^{i,j,(n,m)}} \q^{\bm{\alpha}}:=({\bm{v}^{(i)}_{nm}}^T \phi(\q_i\q_i^T)^T \bm{S}^{(j)} \phi(\q_i\q_i^T) \bm{v}^{(i)}_{nm} - r_{nm}^{(j)})^2,\cr
i=1,\ldots,M,\ j=1,\ldots,N,\ (n,m)\in {E_\mathrm{RDC}}_i.
\end{multline}
We arrive at the following convex program
\begingroup\makeatletter\def\f@size{8}\check@mathfonts
\begin{eqnarray}
\label{general quat optimization 5}
&\ &\min_{\{\tilde{\mathcal{M}}_{\leq 2t,i},\ \bm{y}_i\}_{i\in [1,M]}} \sum_{i=1}^M \sum_{j=1}^N \sum_{(n,m)\in {E_\mathrm{RDC}}_i} \sum_{\bm{\alpha}} {f_{\text{RDC},{\bm \alpha}}^{i,j,(n,m)}} \bm{y}_{\bm{\alpha}} \\
&\quad\text{s.t.} & \tilde{\mathcal{M}}_{\leq 2t,i}\succeq 0,\ i\in[1,M],\cr
&\ & {\bm{y}_i}_{[0,0,0,0]} = 1,\ i\in[1,M],\cr
&\ & \big(\bm{U}_{\leq 2t}   \tilde{\mathcal{M}}_{\leq 2t,i} \bm{U}_{\leq 2t} ^T\big)_{\bm{\alpha} \bm{\beta}} = {\bm{y}_i}_{\bm{\alpha}+\bm{\beta}},\ i\in[1,M],\cr
&\ & \sum_{\bm \alpha} {\bm{l}_{1,i}^{\bm{\gamma}}}_{\bm \alpha} {\bm{y}_i}_{\bm{\alpha}} = \sum_{\bm \alpha} {\bm{l}_{2,i}^{\bm{\gamma}}}_{\bm \alpha} {\bm{y}_i}_{\bm{\alpha}},\ \bm{\gamma}\in\mathbb{N}^3,\ \vert\bm{\gamma}\vert\leq t,\ i\in[2,M]
\end{eqnarray}
\endgroup
for protein structuring from RDC, which is an SDP with $M$ PSD variables of size $\binom{3+t}{3}\times\binom{3+t}{3}$. Again, we typically use $t=4$, where each PSD variable is of size $35 \times 35$. 

\section{Convex relaxation with both NOE and RDC constraints}\label{section:NOE RDC relaxation}
\label{section:quadratic articulated problem}
In the absence of sufficient RDC information, it is important to include NOE restraints along with RDC in order to determine the protein structure. In this section, we describe how to combine RDC-SOS and (\ref{NOE with G relaxed}) to incorporate both types of measurements. Previously in (\ref{NOE with G relaxed}), we relaxed the non-convex equality constraint
\begin{equation}
\G = [\bm{R}_1,\ldots,\bm{R}_M]^T [\bm{R}_1,\ldots,\bm{R}_M]
\end{equation}
into $G\succeq 0$ and forgo using the factors $[\bm{R}_1,\ldots,\bm{R}_M]$ in the factorization of $G$. In this case, only the relaxed version of the relative rotations $\bm{R}_i^T \bm{R}_j$'s are being optimized over, making the formulation in (\ref{NOE with G relaxed}) oblivious to the absolute orientation of each individual rigid unit with respect to the Saupe tensor frame. Therefore in order to incorporate RDC-SOS into (\ref{NOE with G relaxed}), we need to be able to work with each individual rotation matrices directly in (\ref{NOE with G relaxed}). To this end, we use the following different convex relaxation of $\G$:
\begin{equation}
\G \succeq [\bm{R}_1,\ldots,\bm{R}_M]^T [\bm{R}_1,\ldots,\bm{R}_M],
\end{equation}
or equivalently
\begin{equation}
\label{schur complement}
\begin{bmatrix} \bm{G} & \begin{matrix} \R^T_1 \\ \vdots \\ \R^T_M \end{matrix} \\ \begin{matrix} \R_1 & \ldots & \R_M \end{matrix} & \bm{I}_3\end{bmatrix}\succeq 0
\end{equation}
in order to include the variables $\R_1,\ldots,\R_M$ explicitly. Then, the $t=2$ blocks in RDC-SOS can be used to fix the matrices $\R_i$'s in (\ref{schur complement}). The extra information from RDC-SOS on $[\bm{R}_1,\ldots,\bm{R}_M]\in \mathbb{R}^{3\times 3M}$, along with the constraint $\bm{G}\succeq [\R_1,\ldots,\bm{R}_M]^T [\bm{R}_1,\ldots,\bm{R}_M]$ help to concentrate the spectrum of $\G$ into three prominent eigenvalues. In this case, instead of solving the distance geometry problem in a high dimensional space as in (\ref{NOE with G relaxed}), the matrix $\bm{G}$ in RDC-NOE-SOS has much lower rank. This tightens the convex relaxation in (\ref{NOE with G relaxed}) by getting us closer to solving the distance geometry problem in $\mathbb{R}^3$. Here is the resulting convex program RDC-NOE-SOS that incorporates both RDC and NOE:  %Without RDC, the lower bound $[\R_1,\ldots,\R_M]$
\begingroup\makeatletter\def\f@size{8}\check@mathfonts
\begin{eqnarray}
\label{NOE and RDC}
&\ &\min_{\bm{G},\{\tilde{\mathcal{M}}_{\leq 2t,i},\ \bm{y}_i, \R_i\}_{i\in [1,M]}} \lambda f_\text{NOE}(\bm{G}) + \sum_{i=1}^M \sum_{j=1}^N \sum_{(n,m)\in {E_\mathrm{RDC}}_i} \sum_{\bm{\alpha}}  {f_{\text{RDC},\bm{\alpha}}^{i,j,(n,m)}}\bm{y}_{\bm{\alpha}} \\
&\mathrm{s.t.} & \bm{G}_{ii} = \bm{I}_3,\ i\in[1,M],\cr
&\ & \bm{G}\succeq [\R_1,\ldots,\bm{R}_M]^T [\bm{R}_1,\ldots,\bm{R}_M],\label{G lower bound}\\
&\ & \R_i = \phi\bigg(\begin{bmatrix} {\bm{y}_i}_{[2,0,0,0]} & {\bm{y}_i}_{[1,1,0,0]}  & {\bm{y}_i}_{[1,0,1,0]}  & {\bm{y}_i}_{[1,0,0,1]} \\
{\bm{y}_i}_{[1,1,0,0]} & {\bm{y}_i}_{[0,2,0,0]}  & {\bm{y}_i}_{[0,1,1,0]}  & {\bm{y}_i}_{[0,1,0,1]}\\
{\bm{y}_i}_{[1,0,1,0]} & {\bm{y}_i}_{[0,1,1,0]}  & {\bm{y}_i}_{[0,0,2,0]}  & {\bm{y}_i}_{[0,0,1,1]}\\
{\bm{y}_i}_{[1,0,0,1]} & {\bm{y}_i}_{[0,1,0,1]}  & {\bm{y}_i}_{[0,0,1,1]}  & {\bm{y}_i}_{[0,0,0,2]}
\end{bmatrix}\bigg),\ i\in[1,M],\cr
&\ & \tilde{\mathcal{M}}_{\leq 2t,i}\succeq 0,\ i\in[1,M],\cr
&\ & {\bm{y}_i}_{[0,0,0,0]} = 1,\ i\in[1,M],\cr
&\ & \big(\bm{U}_{\leq 2t}   \tilde{\mathcal{M}}_{\leq 2t,i} \bm{U}_{\leq 2t} ^T\big)_{\bm{\alpha} \bm{\beta}} = {\bm{y}_i}_{\bm{\alpha}+\bm{\beta}},\ i\in[1,M],\cr
&\ & \sum_{\bm \alpha} {\bm{l}_{1,i}^{\bm{\gamma}}}_{\bm \alpha} {\bm{y}_i}_{\bm{\alpha}} = \sum_{\bm \alpha} {\bm{l}_{2,i}^{\bm{\gamma}}}_{\bm \alpha} {\bm{y}_i}_{\bm{\alpha}},\ \bm{\gamma}\in\mathbb{N}^3,\ \vert\bm{\gamma}\vert\leq t,\ i\in[2,M].
\end{eqnarray}
\endgroup
The soft penalty $\lambda  f_\text{NOE}(\bm{G})$ is used to enforce the NOE restraints, where the typical choice of $\lambda$ is $10^{-9}$ in order to balance the costs associated with RDC and NOE. Alternatively, since the NOE restraints on the backbone are quite reliable, one can also include them as hard upper and lower bound in RDC-NOE-SOS. We note that for RDC-NOE-SOS, in addition to the $M$ PSD variables $\tilde{\mathcal{M}}_{\leq 2t,i}$ of size $\binom{t+3}{3}\times\binom{t+3}{3}$, we have another PSD variable of size $3(M+1)\times 3(M+1)$ for the constraint (\ref{G lower bound}). In order to take into account the RDC cost which is an 8-th order polynomial in terms of the unit quaternions, we have to choose $t\geq 4$. In our numerical experiments, we always fix $t=4$, because this choice already gives high quality solutions. We note that when using primal-dual interior point method SDP solvers \cite{toh1999sdpt3,sturm1999using,mosek2010mosek} to solve an SDP with $n\times n$ PSD matrix, the typical complexity per iteration is $O(n^{3})$ \cite{krishnan2005interior}. In RDC-NOE-SDP, there are $M$ PSD variables of size $\binom{3+t}{3}\times \binom{3+t}{3}$ and a $3(M+1)\times 3(M+1)$ PSD variable that hosts $G$. This gives $O(M^{3}+M (t^3)^{3})$ per iteration complexity.

\subsection{Rounding}\label{subsec:rounding}
In this section, we describe a \emph{rounding} scheme to extract rotations from the solutions of RDC-SOS and RDC-NOE-SOS. When the matrices $\mathcal{M}_{\leq 2t,i}$'s from the solution to RDC-SOS and RDC-NOE-SOS are rank 1, they correspond to the monomials of a quaternion. If not, two things can happen: (1)  RDC-SOS and RDC-NOE-SOS only manage to find approximate solutions, (2) RDC-SOS and RDC-NOE-SOS exactly recover multiple global minimizers successfully. Denote the solution to RDC-SOS and RDC-NOE-SOS as $\mathcal{M}_{\leq 2t,i}^\star$. The \emph{flat extension theorem} \cite{henrion2005detecting} provides a sufficient condition to check if we fall in the second case. When applied to RDC-SOS and RDC-NOE-SOS, the flat extension theorem says that if
\begin{equation}
\text{rank}(\mathcal{M}_{\leq 2t,i}^\star) = \text{rank}(\mathcal{M}_{\leq 2t-2,i}^\star),
\end{equation}
the global minimizers can be obtained from the Cholesky factorization
\begin{equation}
\mathcal{M}_{\leq 2t,i}^\star = \bm{V}_{\leq t,i} \bm{V}_{\leq t,i}^T
\end{equation}
of $\mathcal{M}_{\leq 2t,i}^\star$. The implementation of the solution extraction algorithm from $\bm{V}_{\leq t,i}$ is described in detail in \cite{henrion2005detecting}. We remind the reader again that $\mathcal{M}_{2t,i}^\star$ can be reconstructed from the variable $\mathcal{M}_{\leq 2t,\text{even},i}^\star$ in RDC-SOS and RDC-NOE-SOS, as mentioned in section \ref{section:size reduction}.

If the condition of the flat extension theorem is not satisfied, there is no guarantee that RDC-SOS and RDC-NOE-SOS return the solution to the un-relaxed problem. However, it is possible to extract an approximate solution based on a heuristic presented below. First, let the rank-1 approximation to $\mathcal{M}_{4,i}^\star\in \mathbb{R}^{10\times 10}$ be
\begin{equation}
\mathcal{M}_{4,i}^\star \approx[\q]^\star_{2,i} {[\q]^\star_{2,i}}^T.
\end{equation}
The vector $[\q]^\star_2\in \mathbb{R}^{10}$ can be seen as an approximation to the degree two monomials of the unit quaternion. After forming a $4\times 4$ matrix $\bm{Q}$ such that
\begin{equation}
{\bm{Q}_i}_{\bm{\alpha}\bm{\beta}}^\star =  {[\q]_{2,i}^\star}_{\bm{\alpha}+\bm{\beta}},\ \vert \bm{\alpha} \vert,\ \vert \bm{\beta} \vert \leq 1,
\end{equation}
the top eigenvector $\q_i^\star$ of $\bm{Q}^\star_i$ is used to find the best rank-1 approximation to $\bm{Q}^\star_i$ and its corresponding rotation $\bm{R}_i^\star = \phi(\q_i^\star {\q_i^\star}^T)$. We note that there is a sign ambiguity when computing $[\q]^\star_{2,i}$, and we choose the sign such that the largest eigenvalue in magnitude of $\bm{Q}_i^\star$ is positive (recall that $\bm{Q}_i$ needs to satisfy rank$(\bm{Q}_i)=1$, $\Tr(\bm{Q})=1$. In particular its only non-zero eigenvalue should be positive).

For the case when RDC-SOS and RDC-NOE-SOS do not give solution that satisfies the condition of the flat extension theorem, the non-convex problem of finding the rotations of the rigid units is not solved exactly. After rounding there is no guarantee that $\bm{R}_i^\star$ orient the rigid units optimally such that the combination of the costs (\ref{RDC minimization}) and (\ref{NOE unrelaxed}) is minimized. In this case, since the pose recovery problem for an articulated structure is an optimization problem on the product of $\mathbb{SO}(3)$ manifolds, we use the manifold optimization toolbox Manopt \cite{boumal2014manopt} to refine $\bm{R}_i^\star$ further in order to obtain a solution with a lower cost. However, since ManOpt only handles unconstrained optimization problems on a Riemanian manifold, we have to use the penalty method to handle the hinge constraint (\ref{hinge constraint}) of the type $h(\bm{R}_i)=0$ by adding a penalty $(\mu/2) \|h(\bm{R}_i)\|_2^2$ with increasing $\mu$. We note that without a good initialization, manifold optimization can easily get stuck in a local minima as it is essentially a gradient descent based approach that descends along the geodesics of a manifold.

\section{Translation Estimation}
\label{section:translation}
In the presence of RDC measurements, the backbone conformation of the full protein can be determined from the calculated $\bm{R}_i$'s, up to a global translation. However, it is usually the case that some of the amino-acid residues contain very few or no RDC's being measured. While RDC-SOS will certainly fail in these situations, using RDC-NOE-SOS is also undesirable. The convex relaxation in (\ref{NOE and RDC}) is typically not tight if some parts of the protein are solely constrained by the NOE. In this case we need to break up the protein and calculate the conformations for selected fragments of the protein backbone. Then we figure out the relative translation between the fragments in order to combine the backbone segments coherently. Using such divide-and-conquer scheme can also speed up the structural calculation process. In this section, we propose a semidefinite relaxation that \emph{jointly} uses NOE restraints between all fragments to piece them together. Let there be $F$ fragments. We denote the coordinate of the $k$-th atom in the $i$-th fragment as $\bm{z}^{(i)}_k$. We note that in this section, the superscript ``$(i)$'' is no longer used as the index for rigid peptide plane or CA-body, but as the index of a fragment composed of multiple amino acid residues. The goal is to find $t_1,\ldots, t_F\in \mathbb{R}^3$ such that
\begin{multline}
\label{interNOE}
({d_{kl}^{\mathrm{low}}})^2 \leq \| \bm{z}^{(i)}_k + \bm{t}_i - (\bm{z}^{(j)}_l + \bm{t}_j)\|_2^2 \leq ({d_{kl}^{\mathrm{up}}})^2,
\end{multline}
where $\quad (k,l)\in E_\mathrm{NOE}$. It should be understood that in this context, $E_\mathrm{NOE}$ only contains the NOE distance restraints between the fragments. The squaring of the constraint is important to obtain a semidefinite relaxation to solve for the pairwise translations. Now let
\begin{eqnarray}
\bm{T} &=&  \begin{bmatrix} \bm{t}_1^T\\ \vdots \\ \bm{t}_F^T \\ \bm{I}_3\end{bmatrix} \begin{bmatrix}\bm{t}_1 &\cdots & \bm{t}_F & \bm{I}_3 \end{bmatrix} \cr
&=& \begin{bmatrix} \bm{t}_1^T \bm{t}_1 & \ldots & \bm{t}_1^T \bm{t}_F & \bm{t}_1^T \\ \vdots & \ddots & \vdots & \vdots \\ \bm{t}_F^T \bm{t}_1 &\ldots & \bm{t}_F^T \bm{t}_F & \bm{t}_F^T \\ \bm{t}_1 & \ldots & \bm{t}_F & \bm{I}_3 \end{bmatrix}\in \mathbb{R}^{(3+F)\times(3+F)}
\end{eqnarray}
where $\bm{T}$ is rank 3 and positive semidefinite. Again, by writing (\ref{interNOE}) in terms of $\bm{T}$ and by relaxing the rank 3 constraint for $\bm{T}$ we can solve for the pairwise translations through the following semidefinite program
\begingroup\makeatletter\def\f@size{8}\check@mathfonts
\begin{alignat}{3}
& &&\min_{\substack{\bm{T}\succeq0,\\ e^{\mathrm{up}}_{kl}\geq 0,\ e^{\mathrm{low}}_{kl}\geq 0}} \sum_{(k,l)\in E_\mathrm{NOE}}  e^{\mathrm{up}}_{kl} +  e^{\mathrm{low}}_{kl} - \gamma\Tr(\bm{T}) \label{tran sync}\\
&\mathrm{s.t.}&& 2(\bm{T}(F+1:F+3,i)-\bm{T}(F+1:F+3,j))^T (\bm{z}^{(i)}_k-\bm{z}^{(j)}_l)\cr
&\ &&   \quad+ \bm{T}(i,i)+\bm{T}(j,j)-2\bm{T}(i,j)+ \|\bm{z}^{(i)}_k-\bm{z}^{(j)}_l\|_2^2\cr
&\ && \quad\quad\leq ({d_{kl}^{\mathrm{up}}})^2 + e^{\mathrm{up}}_{kl},\quad (k,l)\in E_\mathrm{up},\cr
&\ && 2(\bm{T}(F+1:F+3,i)-\bm{T}(F+1:F+3,j))^T(\bm{z}^{(i)}_k-\bm{z}^{(j)}_l)\cr
&\ && \quad \bm{T}(i,i)+\bm{T}(j,j)-2\bm{T}(i,j)+ \|\bm{z}^{(i)}_k-\bm{z}^{(j)}_l\|_2^2 \cr
&\ &&  \quad\quad \geq({d_{kl}^{\mathrm{low}}})^2 - e^{\mathrm{low}}_{kl},\quad (k,l)\in E_\mathrm{low},\cr
&\quad && \bm{T}(F+1:F+3,F+1:F+3) = \bm{I}_3\cr
&\quad && \bm{T}(1:F,1:F)\bm{1} = 0.\nonumber
\end{alignat}
\endgroup
The last constraint is there simply to remove the global translation ambiguity. Instead of using (\ref{interNOE}) as hard constraints to find pairwise translations that satisfy them, we penalize the violation of such bounds through the cost in (\ref{tran sync}). This is necessary because errors in estimating individual fragment coordinates and also ambiguous NOE assignments may cause violations of (\ref{interNOE}). The additional maximum variance unfolding \cite{weinberger2006mvu} type regularization $ -\gamma \Tr(\bm{T})$ prevents the fragments from clustering too tightly by maximizing the spread of the translations \cite{biswas2006snlsdp}. Empirically, we find that a small regularization parameter $\gamma$ between $10^{-3}$ and $10^{-2}$ works well. After obtaining the solution $\bm{T}^\star$, we simply use $\bm{T}^\star(F+1:F+3,1:F)$ as the translations for the fragments.
%This regularization term has a similar effect as using water refinement procedure to improve packing quality in traditional NMR structural calculation.

\begin{figure}[h!]
  \centering
    \includegraphics[width=0.45\textwidth]{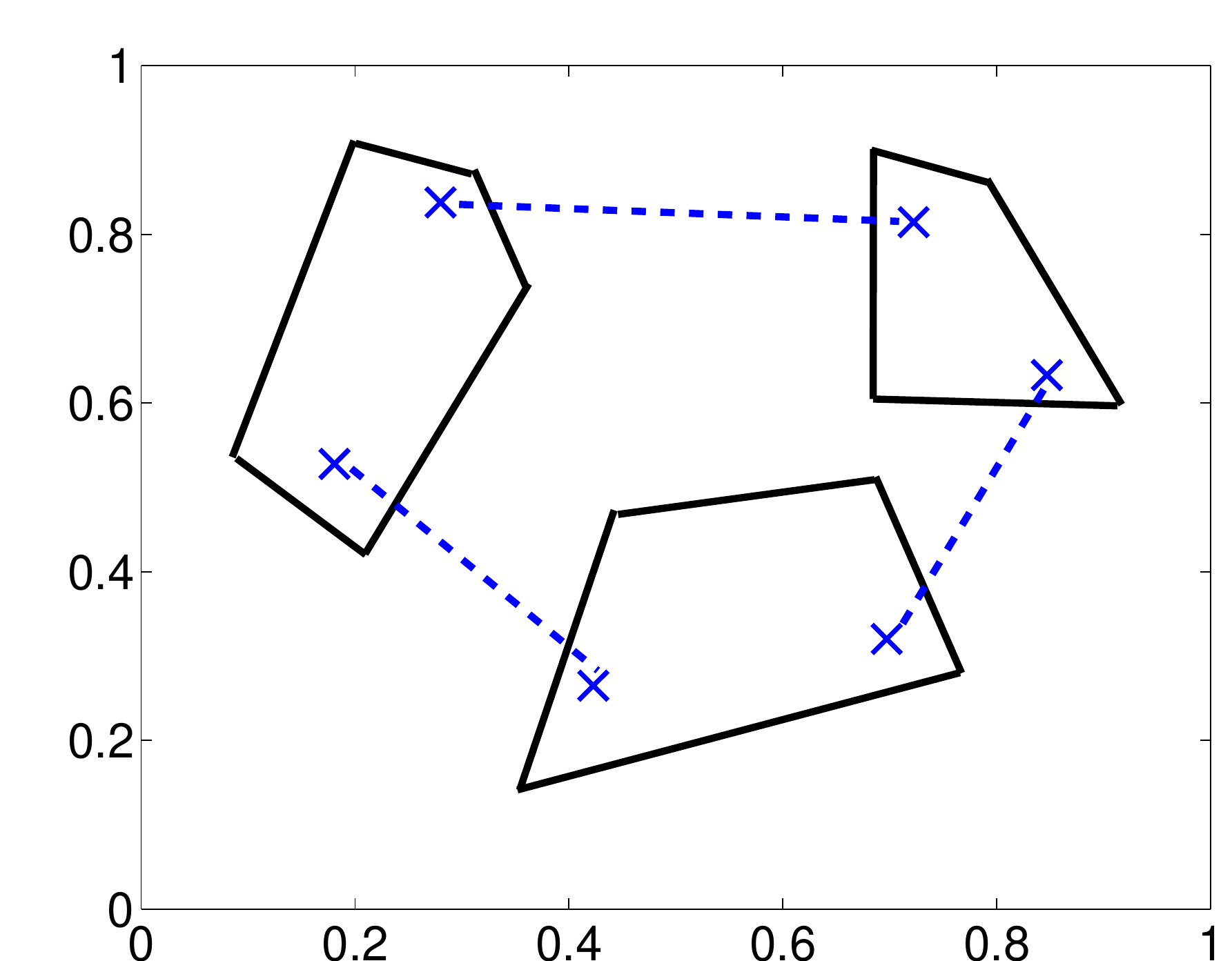}
    \caption{Three fragments in 2D positioned by solving (\ref{tran sync}) using the distance measurements (Blue dotted lines). While it is impossible to determine the translations sequentially with the distance measurement pattern shown here, with the convex program (\ref{tran sync}) the three fragments can be assembled jointly.}\label{figure:translation}
\end{figure}

We conclude this section with a toy example that demonstrates the superiority of joint translation estimation using SDP. For the convenience of illustration, we provide the example in 2D. In order to sequentially assemble the fragments from pairwise distances, it is necessary that there is a pair of fragments where there are at least two distance measurements between them.
This is needed to fix the relative translation between the two fragments with two degrees of freedom.
In the toy example in Figure \ref{figure:translation}, this necessary condition for greedy sequential methods is not satisfied, but even so by solving (\ref{tran sync}) we are able to recover the correct positions of the fragments. This property is quite important, since in practice there are typically only a few NOE restraints between secondary elements of the protein backbone (with the exception of  $\beta$  strands) \cite{mukhopadhyay2014dynafold}.

\section{Numerical experiments}\label{section:numerical}

\subsection{Synthetic data}
\label{section:simulation}
In this section, we present the results of numerical simulations with synthetic data for RDC-SOS and RDC-NOE-SOS. All numerical experiments are run on a Samsung NP940X3G laptop with a Intel(R) Core(TM) i5-4200 2.3GHz CPU and 4 GB of memory. We first describe the noise model in our simulations. Let $\bm{\zeta} = [\bm{\zeta}_1,\ldots,\bm{\zeta}_K]\in\mathbb{R}^{3\times K}$ be the ground truth coordinates. We drop the superscript ``$(i)$'' when denoting the atom coordinate since the membership of an atom to a rigid unit is immaterial here. Now let $E_\mathrm{RDC}$ be the set of atom pairs with RDC measured, and assume that the RDC measurements are generated through
\begin{multline}
\label{RDC noisy}
r_{nm}^{(j)} = {\bm{v}_{nm}}^T  \bm{S}^{(j)} \bm{v}_{nm}+\sigma \epsilon_{nm}^{(j)},\cr
 (n,m)\in {E_\mathrm{RDC}},\ j=1,2,
\end{multline}
where the bond direction $\bm{v}_{nm}$ is related to the coordinates $\bm{\zeta}_n, \bm{\zeta}_m$ through
\begin{equation}
\label{unit vector}
\bm{v}_{nm} = \frac{\bm{\zeta}_n-\bm{\zeta}_m}{\|\bm{\zeta}_n-\bm{\zeta}_m\|_2}.
\end{equation}
We assume $\epsilon_{nm}^{(j)}\sim \mathcal{N}(0,1)$ where $\mathcal{N}(0,1)$ is the standard normal distribution. While it is quite common for different types of atomic pairs with RDC measured at different levels of uncertainty, in this section we assume $r_{nm}$'s are all corrupted by i.i.d. Gaussian noise of same variance $\sigma^2$.

In this simulation study, we use the $\alpha$ helix of the protein ubiquitin (residue 24 - residue 33) to generate synthetic RDC data. The data file for the PDB entry 1D3Z contains RDC datasets measured in two alignment media. From the known PDB structure, we determine the two Saupe tensors $\bm{S}^{(1)}, \bm{S}^{(2)}$ in these alignment media and use them for simulation purposes. We simulate synthetic RDC data using the noise model (\ref{RDC noisy}) where atom pair directions are obtained from the ground truth PDB model. For this simulation we use the pairs $(\text{N}, \text{H}),(\text{C}, \text{CA}),(\text{C}, \text{N})$ from the peptide plane, and $(\text{CA},\text{HA})$ from the CA-body to generate RDCs, as the RDCs associated with these pairs are commonly measured. %Except the RDC between CA and HA, it is possible to obtained the RDC for these pairs of atoms simply via 2D HSQC experiments \cite{hu2006residual}.
In addition to RDC measurements, we also run the simulation with the aid of 16 NOE restraints on the backbone 
for residues 24-33. 
The form of NOE restraints is in terms of hard upper and lower bounds (instead of using them as a penalty term in the cost in RDC-NOE-SOS). To measure the quality of a coordinate estimator $\hat{\bm{\zeta}}$, we use the Root-Mean-Square-Distance (RMSD)
\begin{equation}
\label{RMSD}
\mathrm{RMSD} =  \sqrt{\frac{\|\hat {\bm{\zeta}}- \bm{\zeta}\|_F^2}{K}}
\end{equation}
where $\bm{\zeta}$ is the starting PDB model. We evaluate the RMSD for the atoms CA, CB, C, N, H, O and HA in all amino acids.

\begin{figure}[h!]
  \centering
\centering
    \includegraphics[width=0.4\textwidth]{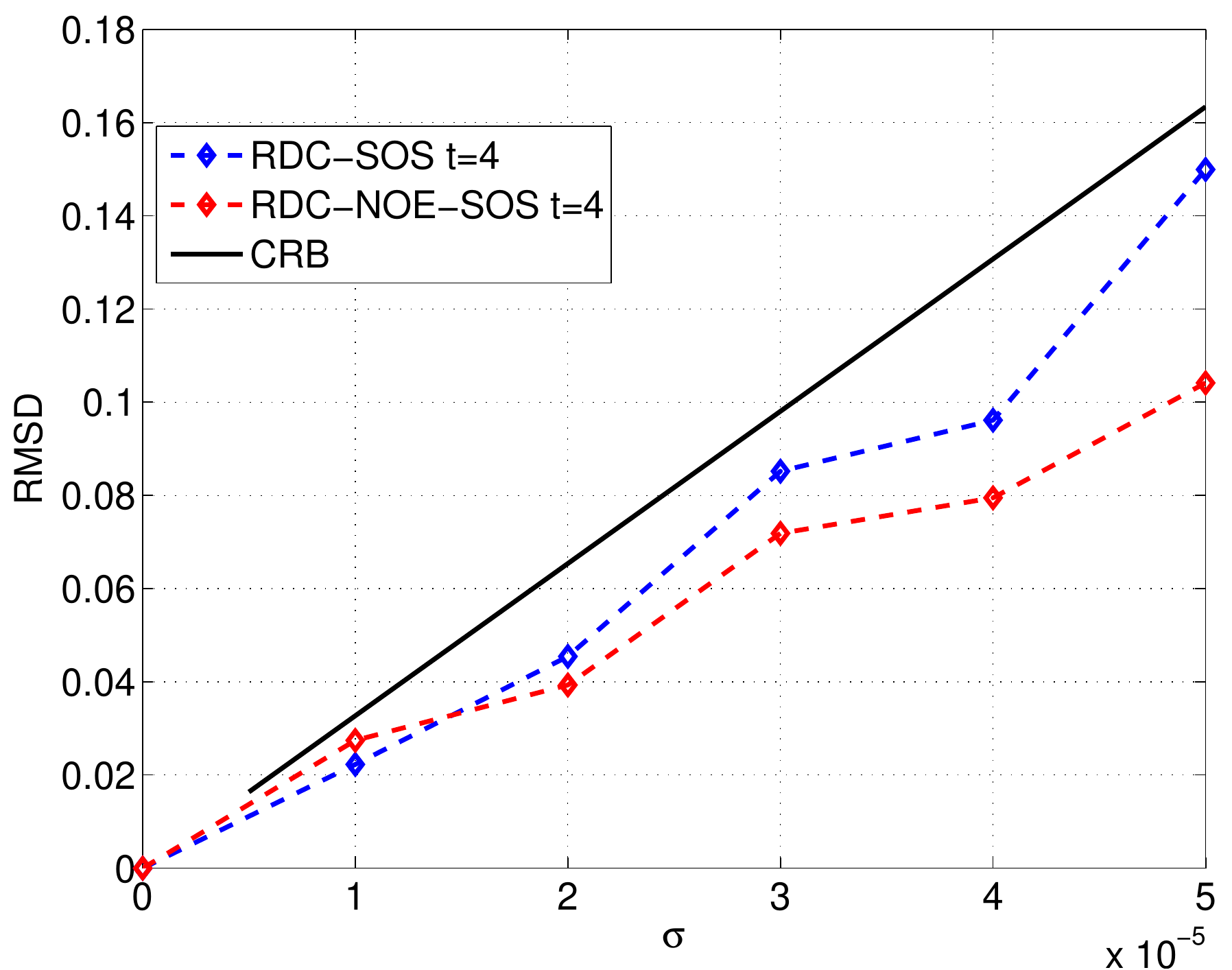}
\caption{Comparison between running RDC-SOS and RDC-NOE-SOS with $t=4$. Here we do not use manifold optimization to further refine the solution.}\label{figure:CRB compare}
\end{figure}

We present the simulation results in Figure \ref{figure:CRB compare}. We simulate RDC noise with $\sigma \in [0, 5\text{e-5}]$. Every data point is averaged over 40 noise realizations of RDC. When there is no noise, RDC-SOS and RDC-NOE-SOS exactly recover the rotations with $t=4$. This is a property that simulated annealing based methods do not enjoy, as even without noise these methods can still suffer from local minima. In this simulation,  $\mathcal{M}_{2t,\text{even},i}^\star$ returned with $t=4$ are rank-1 to $10^{-2}$ precision most of the time, therefore we do not use manifold optimization to further refine the solution. We further compare our results against the Cram\'er-Rao lower bound. The CRB (formula given in Section  \ref{section:compute CRB}) provides an information-theoretic lower bound for the least possible variance that can be achieved by any unbiased coordinate estimator. With RDC-SOS and RDC-NOE-SOS we are able to obtain RMSD lower than the CRB. Since RDC-SOS and RDC-NOE-SOS can produce biased coordinate estimators, their error can be lower than the CRB. Here we remark that we slightly abused terminology by referring to the normalized RDC as RDC, where the un-normalized RDC is defined in (\ref{RDC full}). We emphasize that when $\sigma = \text{5e-5}$, the magnitude of noise on the un-normalized RDC is rather large. For example, since the dipolar coupling constant for the N-H RDC is about 23 kHz, when $\sigma = \text{5e-5}$ the actual noise is 1.15 Hz. This is larger than the typical experimental uncertainty of N-H RDC ($<$0.5 Hz) \cite{hu2006residual}.

\subsection{Experimental data}
\label{section:Real data}

In this section, we present results on the analysis of experimental RDC data obtained in two alignment media for ubiquitin. We also provide a comparison of our methods with the molecular fragment replacement (MFR) method proposed in \cite{bax2001rdc} using the full ubiquitin sequence with 76 amino acids and about 500 backbone atoms. We first give a brief introduction to the MFR method. MFR is an RDC-based method that determines the structure of a protein through finding homologous structures in the PDB for short fragments of the protein. For a short fragment, candidate structures from the PDB are used to construct the coordinates in (\ref{rdc constraint}). Then a least-squares procedure detailed in the appendix is used to obtain the Saupe tensor based on the experimentally measured RDC and the candidate structure. If a PDB candidate structure gives a low residual in the least-squares fitting, it will be deemed a structure similar to the protein fragment under inspection. Other experimental information such as chemical shifts can also be compared to the information recorded in the database to find a similar structure. The homologous structures for short fragments of the protein are then merged and simulated annealing is applied to further refine the structure based on the RDC measurements. In this numerical study, we start simulated annealing with temperature of 600 K and cool down to 0 K in 30000 steps. For a fair comparison between MFR and our proposed methods, we do not use chemical shift information for the MFR procedure but only RDC and NOE.

We only consider the peptide planes and CA-bodies coming from the first 70 amino acids since the last 6 residues are highly flexible and do not contribute to rigid constraints. We solve for the structure using RDC measurements from the bonds $(\text{C}, \text{N}), (\text{N}, \text{H}), (\text{CA},\text{HA})$ in two alignment media. Here we do not use $(\text{C}, \text{CA})$ RDCs as in the previous subsection in order to demonstrate the usefulness of the proposed methods when there are less data. We also examine the situation when we are supplemented with 187 experimentally reported backbone NOE's. We use RDC-SOS and RDC-NOE-SOS with $t=4$ to solve the backbone structure of six ubiquitin fragments, each containing 12-13 residues on average. We split the fragments at amino-acid sites where there are too few or no RDC measurements. The results are summarized in Table \ref{table:main results}. When using only RDC, it is more difficult to determine the backbone structure near the starting and end point of a fragment since RDC measurements are generally sparser in those regions. In this situation, having extra distance constraints may help.

\begin{figure}[h!]
  \centering
    \begin{subfigure}[b]{.43\linewidth}\centering\includegraphics[width=1\textwidth]{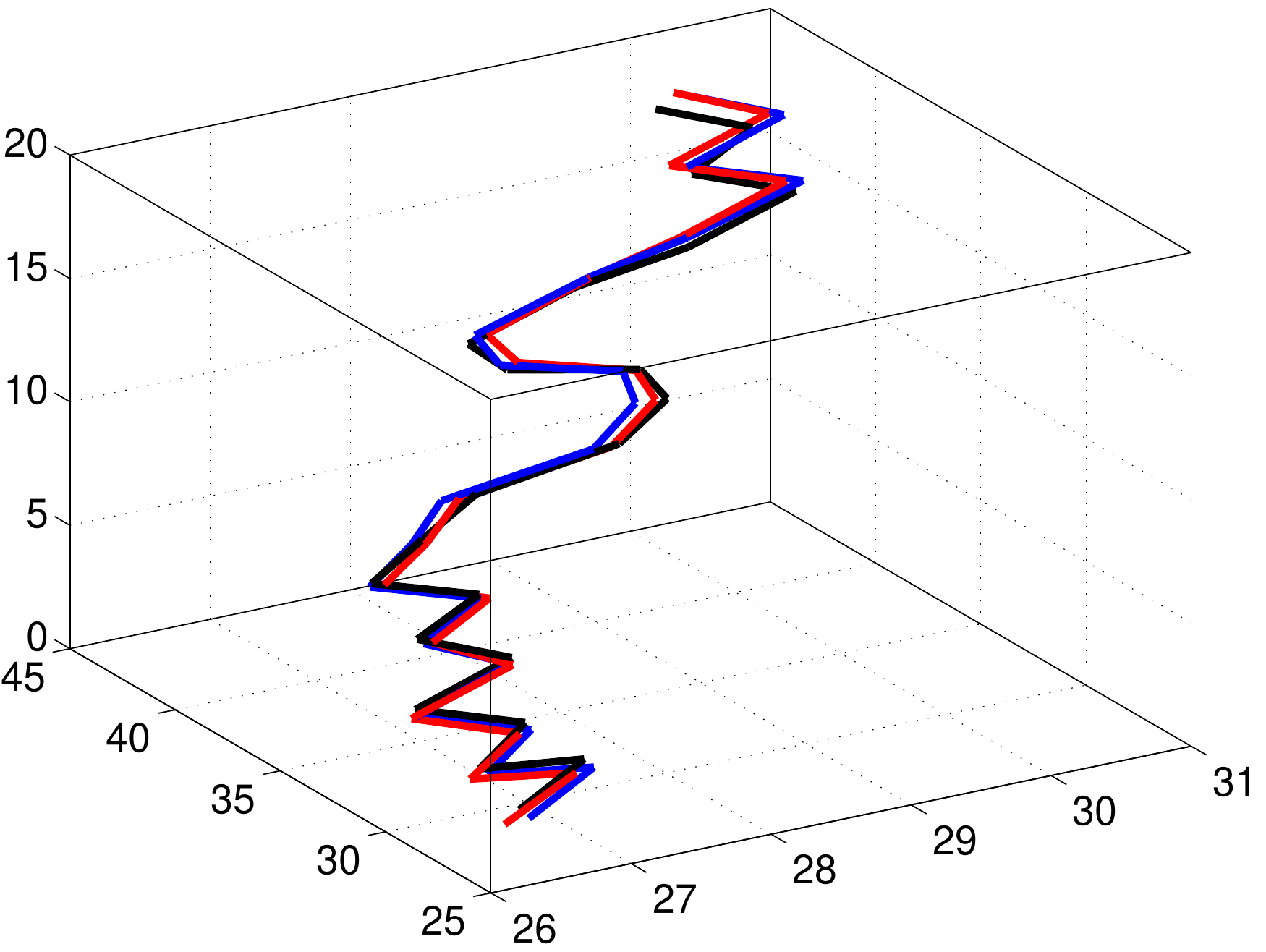}\caption{\centering}\end{subfigure}
    \begin{subfigure}[b]{.43\linewidth}\centering\includegraphics[width=1\textwidth]{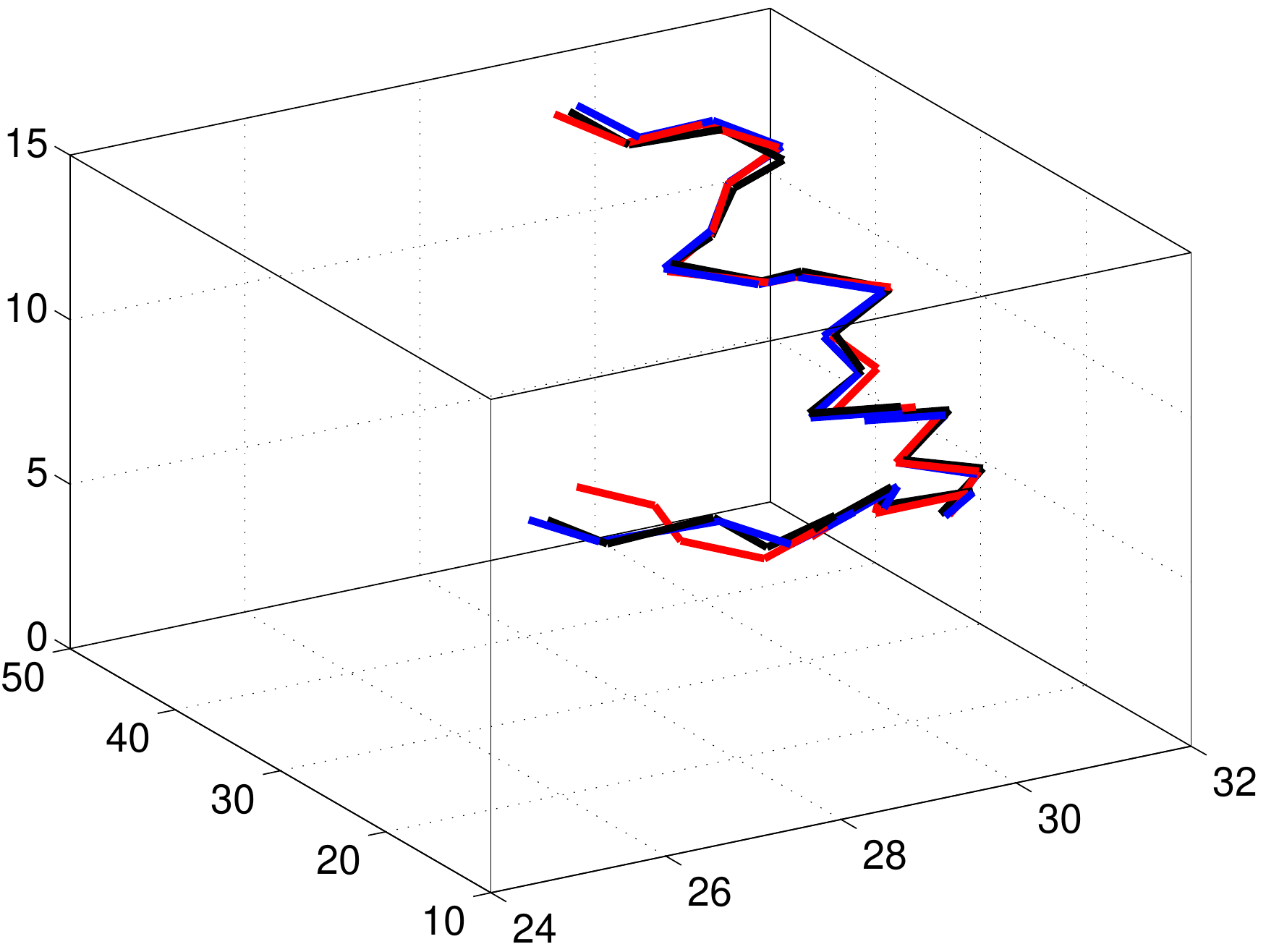}\caption{\centering}\end{subfigure}
    \begin{subfigure}[b]{.43\linewidth}\centering\includegraphics[width=1\textwidth]{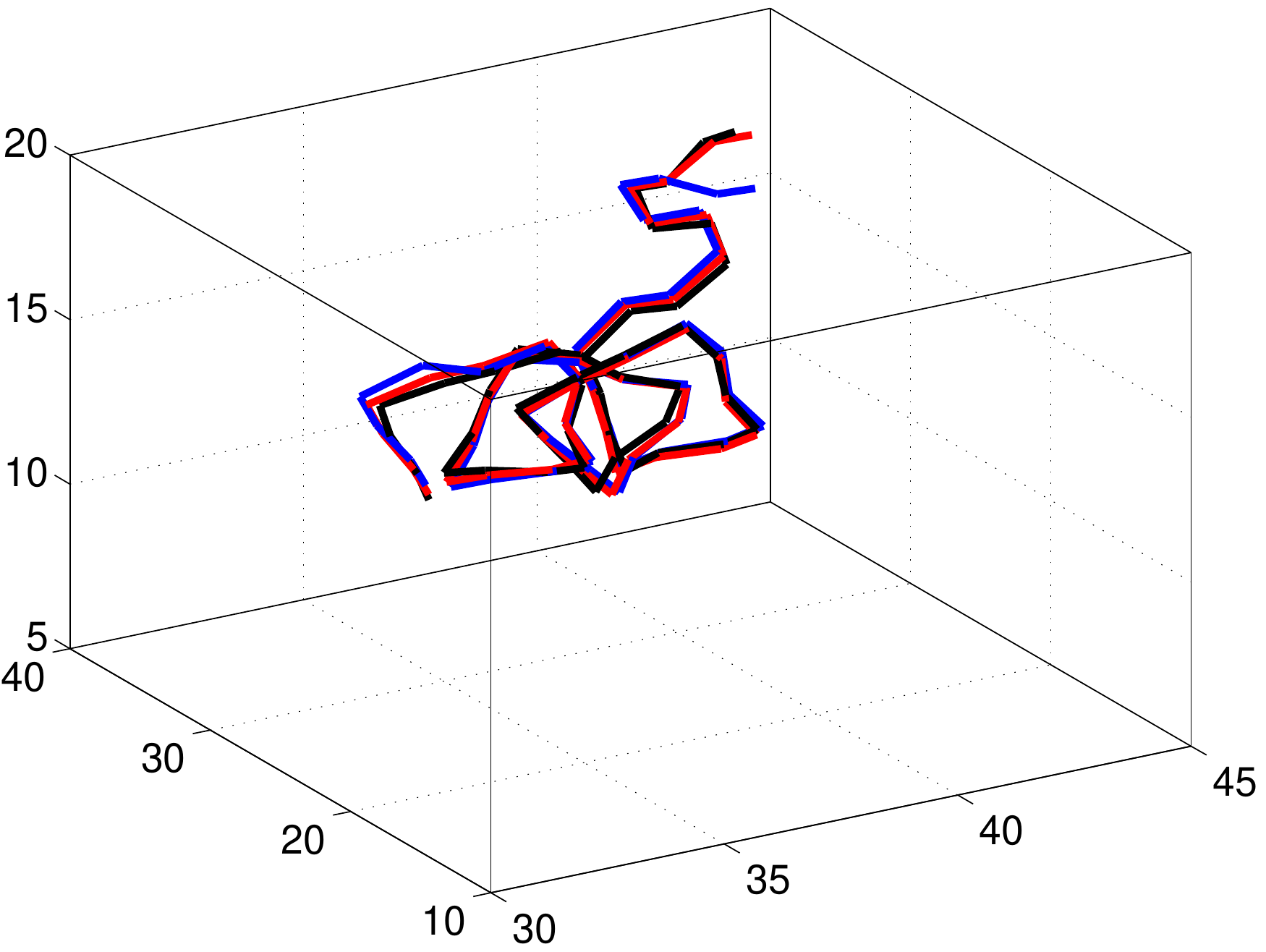}\caption{\centering}\end{subfigure}
    \begin{subfigure}[b]{.43\linewidth}\centering\includegraphics[width=1\textwidth]{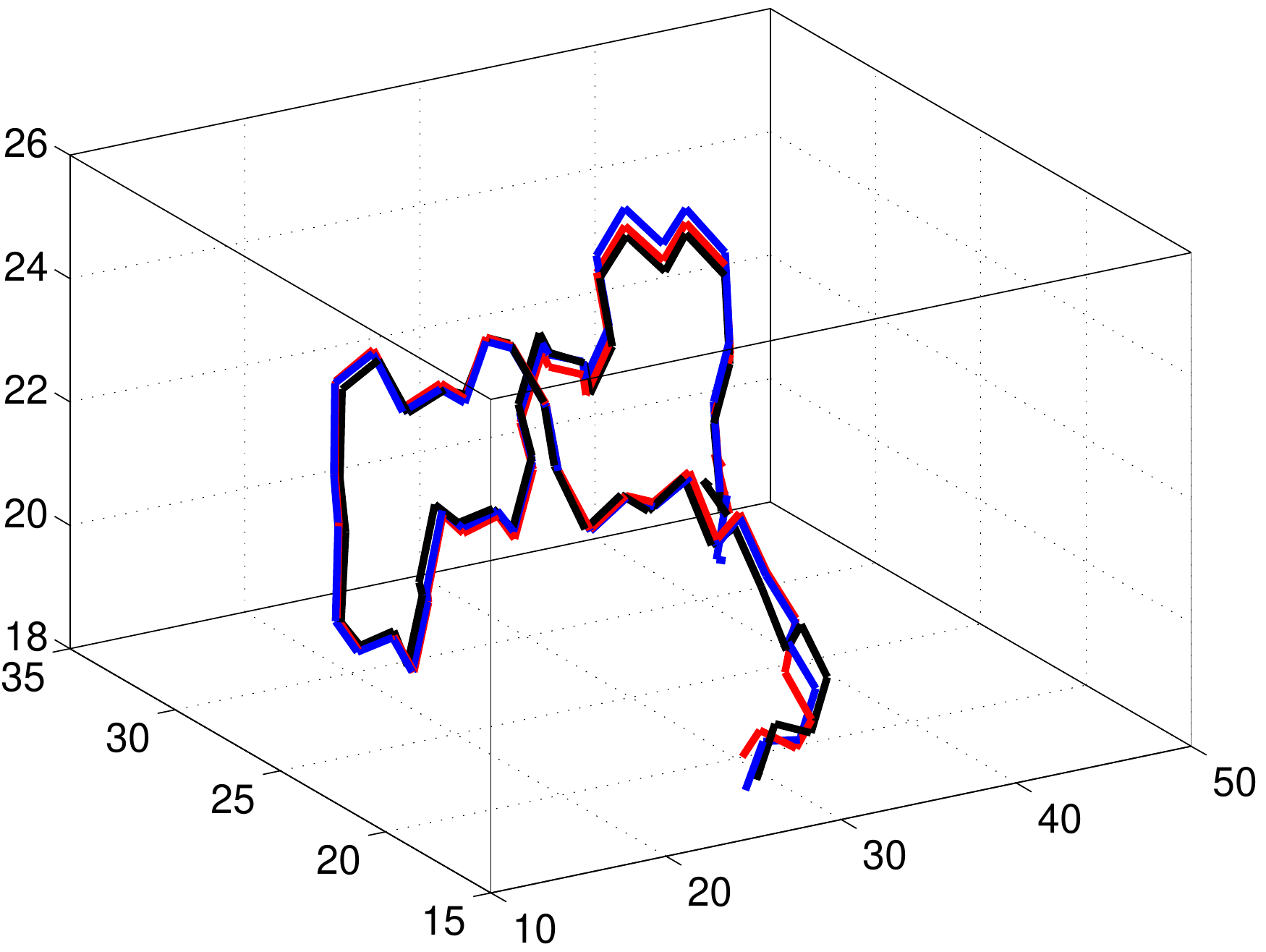}\caption{\centering}\end{subfigure}
    \begin{subfigure}[b]{.43\linewidth}\centering\includegraphics[width=1\textwidth]{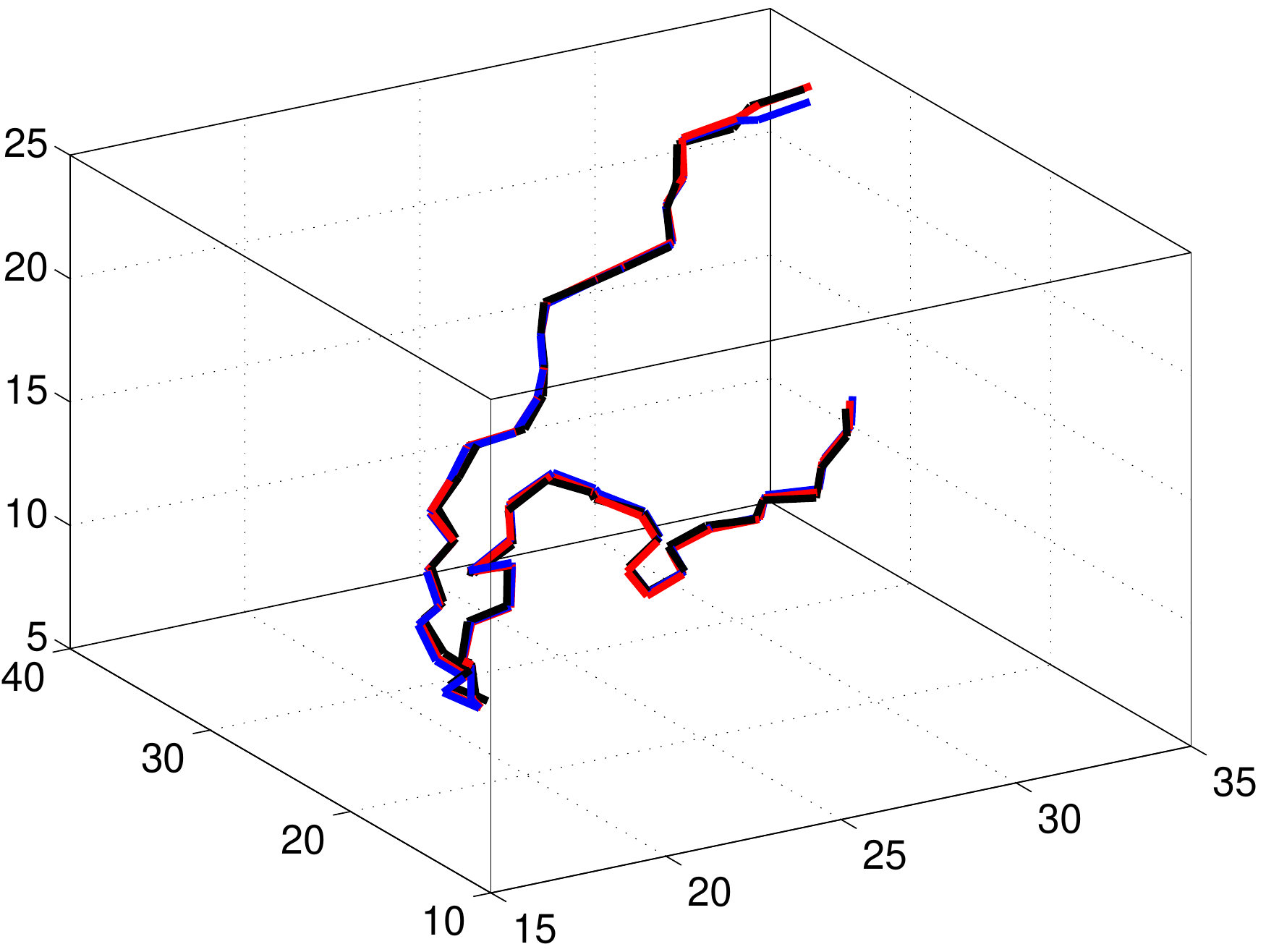}\caption{\centering}\end{subfigure}
    \begin{subfigure}[b]{.43\linewidth}\centering\includegraphics[width=1\textwidth]{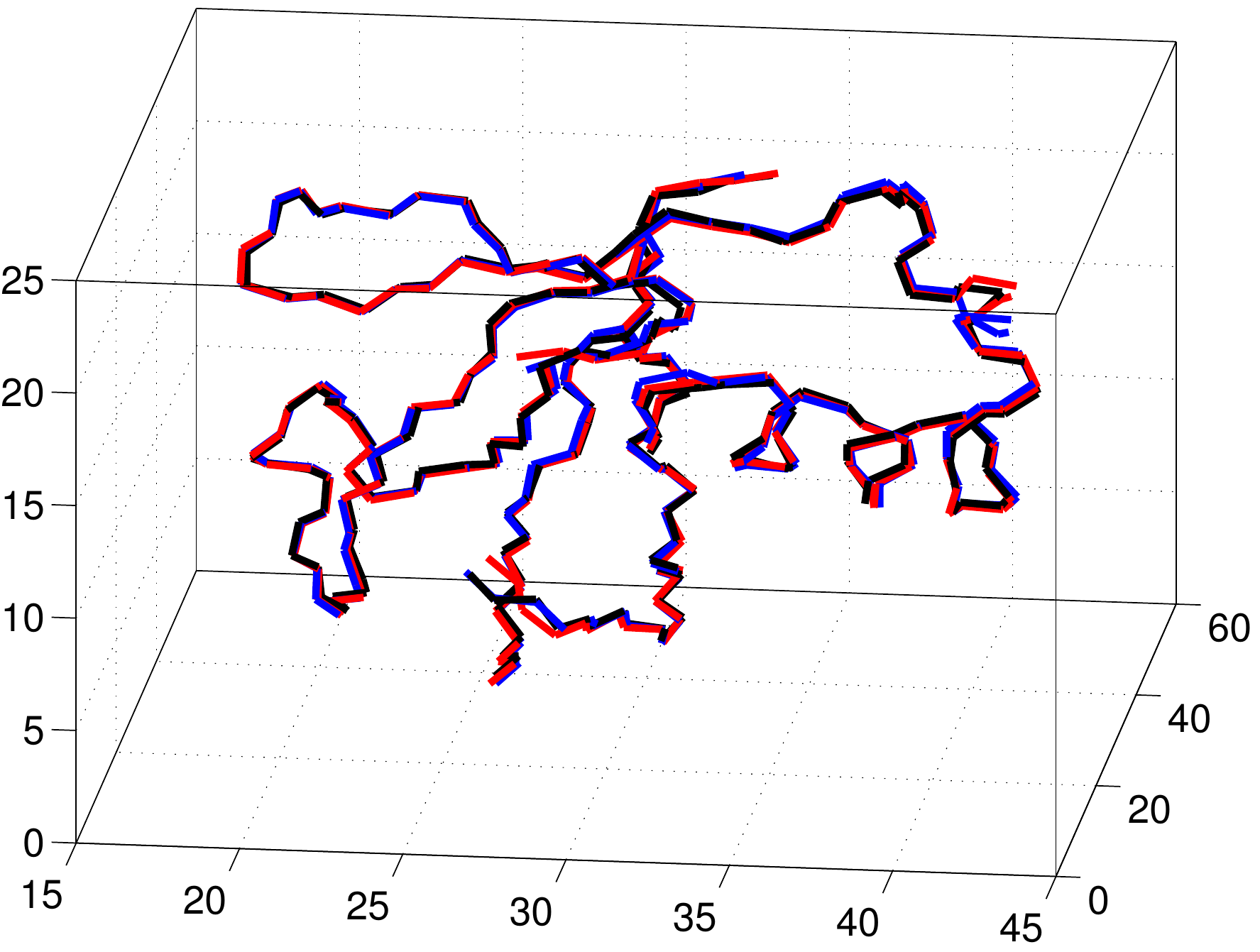}\caption{\centering}\end{subfigure}
    \caption{The trace of protein backbone drawn using N, CA and C. (a),(b),(c),(d),(e): Fragments 1,2,3,4 and 5 of ubiquitin defined in Table \ref{table:main results}. The black, blue and red curves come from the X-ray model 1UBQ, RDC-SOS solution and RDC-NOE-SOS respectively. (f): Full backbone structure obtained from assembling the five ubiquitin fragments.}\label{figure:fragments backbone}
\end{figure}

In terms of accuracy, due to the additional distance restraints, RDC-NOE-SOS outperforms RDC-SOS, except for fragment 2. The lack of constraints on residue 9 and 18 causes RDC-SOS and RDC-NOE-SOS to give different solutions with the same cost. The average RMSD of the fragments are 0.47 \AA \ and 0.39 \AA\ for RDC-SOS and RDC-NOE-SOS respectively when comparing with the X-ray structure 1UBQ \cite{vijay19871ubq}. To provide a different perspective, we also compare the results from our method with the high resolution NMR structure 1D3Z \cite{cornilescu19981d3z}. When we combine the fragments using (\ref{tran sync}), the conformation errors of the whole protein backbone obtained from fragments determined by RDC-SOS and RDC-NOE-SOS are 1.05 (1.00) \AA \ and 0.86 (0.80) \AA \ RMSD respectively when comparing to 1UBQ (1D3Z). Figure \ref{figure:fragments backbone} further compares the backbone traces obtained from our proposed methods and the X-ray structure. Our results are competitive when comparing to MFR, which gives average fragment RMSD of 0.54 \AA\ and overall RMSD of 0.87 \AA. Since RDC-SOS only involves $M$ PSD variables of size $35\times 35$ when $t=4$, whereas RDC-NOE-SOS involves another PSD variable of size $3M\times 3M$, the running time of RDC-SOS is faster than RDC-NOE-SOS.

\begin{table}[H]
\centering % used for centering table
\scalebox{0.8}{
\begin{tabular}{c c c c c c c} % centered columns (4 columns)
\hline\hline %inserts double horizontal lines
Fragment No.&  & 1 & 2 & 3 & 4 & 5\\
\hline\hline % inserts single horizontal line
Residue No. &  & 1-7 & 9-18 & 22-36 & 37-53 & 54-70\\
\hline
\multirow{2}{*}{\begin{tabular}[c]{@{}l@{}}RMSD (\AA)\\ 1UBQ\end{tabular}}  & RDC-SOS  & 0.36 & 0.34 & 0.51 & 0.56 & 0.57  \\
 & RDC-NOE-SOS  &  0.37 & 0.51 & 0.31 & 0.51 & 0.25 \\
 & MFR  & 0.42 & 0.51 & 0.45 & 0.78 & 0.52\\[1ex]
\hline
\multirow{2}{*}{\begin{tabular}[c]{@{}l@{}}RMSD (\AA)\\ 1D3Z\end{tabular}} & RDC-SOS  & 0.33 & 0.25 & 0.46 & 0.51 & 0.54  \\
 & RDC-NOE-SOS  &  0.26 & 0.49 & 0.20 & 0.50 & 0.17 \\
 & MFR  & 0.40 & 0.46 & 0.42 & 0.71 & 0.44\\[1ex]
\hline
Time (s) & RDC-SOS   &  14 & 18 & 30 & 35 & 33 \\
     & RDC-NOE-SOS  & 20 & 22 & 80 & 111 & 114\\
     & MFR & \multicolumn{3}{l}{1560 (all 5 fragments)} & & \\[1ex]
\hline
\end{tabular}} \\ [1ex]
\caption{Results of computing the structure of five ubiquitin fragments using RDC-SDP, RDC-NOE-SDP and MFR from experimental data. We compare with both the X-ray structure 1UBQ and the high resolution NMR structure 1D3Z. For MFR we only report the total running time for calculating the entire backbone.}\label{table:main results}% title of Table
\end{table}

%
%DC we may want a bigger copy of this is supplementaty material ?

%
%\begin{figure}[h!]
%  \centering
%    \includegraphics[width=0.6\textwidth]{figures/frag3.eps}
%    \caption{An example of a fragment with 0.47 \AA \  RMSD.}\label{figure:fragments backbone}
%\end{figure}

\section{Conclusion}
We presented two novel convex relaxations RDC-SOS and RDC-NOE-SOS to calculate the protein backbone conformation from both RDC and NOE measurements.
In simulations, our methods exactly recover the protein structure when there is no noise, whereas simulated annealing based methods can still get trapped at local minima even when the data is clean. We demonstrate the robustness of our algorithm in simulation by showing that in the presence of noise, the error of our solution attains the CRB. We further demonstrated the success of our methods by obtaining a backbone structure of 1 \AA\ resolution for ubiquitin using experimental data. Both proposed methods are fast in practice, in particular RDC-SOS can determine a protein fragment of typical size in less than half a minute. In comparison, the running time of current methods such as MFR, RDC-Analytics and REDCRAFT range from tens of minutes to two hours. This property of our algorithm can be useful when iterating between estimating resonance or NOE assignments and structural calculation \cite{guntert2004cyana}. 

There are a few remaining problems we would like to address in future works. In this paper, we only use the SOS hierarchy with $t=4$ which is sufficient for the numerical experiments considered. We would like to investigate the utility of $t>4$ in the SOS hierarchy for sparse RDC data, especially with only one alignment media.

At this point, both RDC-SOS and RDC-NOE-SOS can only compute the structure of the protein backbone but not the protein side-chains. RDC measurements on side chains are complicated by the existence of rotamer states and only a few recent analyses are able to address this issue \cite{li2015side}. 	
%Since there are few RDC measurements on the protein side-chains and the side-chain NOE restraints are less complete, calculating the structure of the side-chains can be challenging.
%
On the other hand, a major obstacle of obtaining complete NOE restraints for the protein side-chains is the ambiguity in NOE assignment, especially for larger systems. We hope to extend our proposed methods to help detecting the correct NOE assignments for the side-chains, through providing a high quality backbone conformation for assignment validation.

Currently, our method requires sufficient NOE restraints between the fragments when combining the fragments together using the convex program (\ref{tran sync}). Hcowever, as noted in \cite{yershova2011rdc}, there can be very few NOE restraints between the secondary structural elements. We observed such a situation when applying our algorithm to the protein DinI (PDB ID: 1GHH). While all the fragments in DinI can be determined by our proposed method to within 1 \AA\ resolution, our method failed to assemble the fragments together due to the lack of inter-fragments NOE. We hope to solve this issue in the future by including database derived restraints. For example, torsion angle restraints can be derived from chemical shifts of backbone atoms using TALOS \cite{shen2009talos+}. Furthermore, side-chain rotamer library \cite{lovell2000penultimate} can be used to model protein side-chains, which can in turn provide additional NOE restraints arising from the side-chains.

\section{Acknowledgements}
The authors would like to thank James Saunderson for discussions related to unit quaternion parameterization for optimization problems on $\mathbb{SO}(3)$. The authors are grateful to Jo\~{a}o M. Pereira, Roy R. Lederman and Yutong Chen for discussions regarding this problem, to Nicolas Boumal for the discussion on manifold optimization and proof-reading an earlier version of this manuscript. The authors also want to thank Richard Harris and Roberto Tejero for assisting with interpreting and reading NMR restraint files. The research of AS was partially supported by award R01GM090200 from the NIGMS, by awards FA9550-12-1-0317 from AFOSR, by the Simons Foundation investigator award and the Simons Foundation Collaboration on Algorithms and Geometry, and the Moore Foundation Data Driven Discovery Investigator award.

\section{Appendix}\label{section:appendix}
\subsection{The residual dipolar coupling and Saupe tensor}
We give here a brief introduction to RDC and the Saupe tensor, while a detailed exposition can be found in \cite{tolman2006rdc} for example. Let $\bm{v}_{nm}$ be the unit vector denoting the direction of the bond between nuclei $n$ and $m$. Let $b$ be the unit vector denoting the direction of the magnetic field. The RDC $D_{nm}$ due to the interaction between nuclei $n$ and $m$ is
\begin{equation}
D_{nm} = D_{nm}^\text{max} \left \langle \frac{3 (\bm{b}^T \bm{v}_{nm})^2 -1}{2} \right \rangle_{t,e}.
\end{equation}
$D_{nm}^\text{max}$ is a constant depending on the gyromagnetic ratios $\gamma_n,\gamma_m$ of the two nuclei, the bond length $r_{nm}$, and the Planck's constant $h$ as
\begin{equation}
D_{nm}^\text{max} = -\frac{\gamma_n \gamma_m h}{2 \pi^2 r_{nm}^3},
\end{equation}
and $\langle\ \cdot \rangle_{t,e}$ denotes the ensemble and time averaging operator.
As presented, RDC depends on the relative angle between the magnetic field and the bond. By extension of terminology, we refer to the normalized RDC
\begin{equation}
r_{nm} =D_{nm}/D_{nm}^\text{max}
\end{equation}
as simply the RDC.

It is conventional to interpret the RDC measurement in the molecular frame. More precisely, we treat the molecule as being static in some coordinate system, and the magnetic field direction being a time and sample varying vector. In this case the RDC becomes
\begin{equation}
\label{RDC full}
D_{nm} = D_{nm}^\text{max} \bm{v}_{nm}^T \bm{S} \bm{v}_{nm},
\end{equation}
where the Saupe tensor $S$ is defined as
\begin{equation}
\label{Saupe tensor}
\bm{S} = \frac{1}{2}(3 \bm{B} - \bm{I}_3),\qquad \bm{B} =  \left \langle \bm{bb}^T \right \rangle_{t,e}.
\end{equation}
We note that $\bm{S}$ is symmetric and $\Tr(\bm{S}) = 0$. In order to use RDC for structural refinement of a protein, $\bm{S}$ is usually first determined from a known structure (known $\bm{v}_{nm}$) that is similar to the protein.

We now detail a classical way of obtaining the Saupe tensor from a known template structure \cite{losonczi1999order}. Using the fact that $\bm{S}$ is symmetric and $\Tr(\bm{S}) = 0$, eq. (\ref{RDC full}) can be rewritten as
\begin{multline}
\label{RDC for bond nm}
r_{nm} = ({\bm{v}_{nm}}_2^2-{\bm{v}_{nm}}_1^2) \bm{S}(2,2) + ({\bm{v}_{nm}}_3^2-{\bm{v}_{nm}}_1^2) \bm{S}(3,3) \cr
+ 2 {\bm{v}_{nm}}_1 {\bm{v}_{nm}}_2 \bm{S}(1,2) + 2 {\bm{v}_{nm}}_1 {\bm{v}_{nm}}_3 \bm{S}(1,3) \cr
+ 2 {\bm{v}_{nm}}_2 {\bm{v}_{nm}}_3 \bm{S}(2,3)
\end{multline}
where ${\bm{v}_{nm}}_i$, $i=x,y,z$ are the different components of $\bm{v}_{nm}$ in the molecular frame. When there are $L$ RDC measurements, eq. (\ref{RDC for bond nm}) results in $L$ linear equations in five unknowns ($\bm{S}(2,2),\bm{S}(3,3),\bm{S}(1,2),\bm{S}(1,3)$ and $\bm{S}(2,3)$), that can be written in matrix form as
\begin{multline}
\label{ls prestegard}
\bm{A} s = \bm{r},\quad \bm{s} =  \begin{bmatrix} \bm{S}(2,2)\\ \bm{S}(3,3)\\ \bm{S}(1,2)\\ \bm{S}(1,3)\\ \bm{S}(2,3)  \end{bmatrix} \in \mathbb{R}^5, \quad \bm{r} =\begin{bmatrix} r_{n_1m_1}\\ \vdots \\r_{n_L m_L} \end{bmatrix} \in \mathbb{R}^M
\end{multline}
and $\bm{A}\in \mathbb{R}^{L\times 5}$. An ordinary least squares procedure can be used to estimate $s$ if $\bm{A}$ has full rank. This is also referred to as the SVD procedure in \cite{losonczi1999order}.

\subsection{Sum-of-squares relaxation}
\label{section:SOS}
In this section, we explain why the convex relaxation presented in Section \ref{section:RDC relaxation} is coined SOS. The polynomial optimization problem
\begin{equation}
\label{primal nonneg}
p_1 = \min_{\bm{x}\in\mathbb{R}^n} f(\bm{x})\quad \text{s.t.}\ h(\bm{x})=0,
\end{equation}
where $f(\bm{x}),h(\bm{x})$ are polynomial functions, can be expressed equivalently as
\begin{equation}
\max_{d} d\quad \text{s.t.}\ f(\bm{x})-d\geq 0\ \ \ \text{on}\ h(\bm{x})=0.
\end{equation}
This is equivalent to
\begin{equation}
d_1 = \max_{d,t_{\bm{\alpha}}} d\quad \text{s.t.}\ f(\bm{x})-d+(\sum_{\bm{\alpha}}t_{\bm{\alpha}}\bm{x}^{\bm{\alpha}}) h(\bm{x})\geq 0\ \ \ \forall\ \bm{x}
\end{equation}
\cite[Chapter 3]{blekherman2011semidefinite}, which is actually the dual problem to (\ref{primal nonneg}). However, due to the NP-hardness in testing the non-negativity of a polynomial \cite{blekherman2011semidefinite}, we further restrict the search space from the set of non-negative polynomials to the set of SOS polynomials: 
\begin{equation}
\label{SOS dual}
d_2 = \max_{d,t_{\bm{\alpha}}} d\quad \text{s.t.}\ f(\bm{x})-d+(\sum_{\bm{\alpha}}t_{\bm{\alpha}}\bm{x}^{\bm{\alpha}}) h(\bm{x})\ \text{is SOS}.
\end{equation}
This results in a standard SDP 
\begin{equation}
\label{dual SDP}
\max_{d,\bm{P}\succeq 0,t_{\bm{\alpha}}} d\quad \text{s.t.}\ f(\bm{x})-d+(\sum_{\bm{\alpha}}t_{\bm{\alpha}}\bm{x}^{\bm{\alpha}}) h(\bm{x}) = [\bm{x}]^T_{t} \bm{P} [\bm{x}]_{t}
\end{equation}
for some specific choices of $t$. Since $p_1 = d_1\geq d_2$, solving (\ref{dual SDP}) provides a lower bound to (\ref{primal nonneg}). Indeed, the dual of (\ref{dual SDP}) is exactly the type of convex relaxations presented in Section \ref{section:RDC relaxation} for optimization problems of the form (\ref{primal nonneg}).

\subsection{Cram\'er-Rao lower bound}
\label{section:Cramer-Rao}
In this section, we introduce a classical tool from statistics, the Cram\'er-Rao bound (CRB)
\cite{casella2002statistical}, to give perspective on the lowest possible error any unbiased estimator can achieve when estimating coordinates from noisy RDC measurements. We first describe the CRB for general point estimators. Let $\bm{\theta}\in \mathbb{R}^n$ be a multidimensional parameter which is to be estimated from measurements $\bm{x} \in \mathbb{R}^{m}$. Suppose $\bm{x}$ is generated from the distribution $p(\bm{x}\vert \bm{\theta})$. The Fisher information matrix (FIM) is defined as the $n\times n$ matrix
\begin{equation}
\label{FIM}
\bm{I}(\bm{\theta}) = \mathbb{E}[(\nabla_{\bm{\theta}} \ln p(\bm{x} \vert \bm{\theta})) (\nabla_{\bm{\theta}} \ln p( \bm{x} \vert \bm{\theta}))^T]
\end{equation}
where expectation is taken with respect to the distribution $p(\bm{x}\vert \bm{\theta})$ and the gradient $\nabla_{\bm{\theta}}$ is taken with respect to $\bm{\theta}$. For any unbiased estimator $\hat{\bm{\theta}}$ of $\bm{\theta}$, that is $\mathbb{E}(\hat{\bm{\theta}}) = \bm{\theta}$, the following relationship holds:
\begin{equation}
\label{CRB}
\mathbb{E}[(\hat{\bm{\theta}} - \bm{\theta})(\hat{\bm{\theta}} - \bm{\theta})^T] \succeq \bm{I}(\bm{\theta})^{-1}
\end{equation}
if $\bm{I}(\bm{\theta})$ is invertible. Therefore the total variance of the estimator $\hat{\bm{\theta}}$ is lower bounded by $\Tr (\bm{I}(\bm{\theta})^{-1})$. We remark that for an unbiased estimator, its variance and the mean-squared error are the same, therefore we often use these terms interchangeably. 

We also introduce the CRB in the case when $\bm{\theta}$ and $\hat{\bm{\theta}}$ are constrained to be in the set $\{\bm{\theta}\vert \ f(\bm{\theta})= 0\}$ where $f:\mathbb{R}^n \rightarrow \mathbb{R}^k$ \cite{stoica1998crb}. Let $\bm{Df}(\bm{\theta})\in\mathbb{R}^{k\times n}$ be the gradient matrix of $f$ at $\bm{\theta}$ with full row rank, and $\bm{Q}\in \mathbb{R}^{n\times (n-k)}$ be a set of orthonormal vectors satisfying
\begin{equation}
\bm{Df}(\bm{\theta})\bm{Q} = 0
\end{equation}
i.e. $\bm{Q}$ is an orthonormal basis of the null space of $\bm{Df}(\bm{\theta})$. In this case, for any unbiased estimator $\hat{\bm{\theta}}$ satisfying $f(\hat{\bm{\theta}}) = 0$, the CRB is then
\begin{equation}
\label{constrained CRB}
\mathbb{E}[(\hat{\bm{\theta}} - \bm{\theta})(\hat{\bm{\theta}} - \bm{\theta})^T] \succeq \bm{Q}(\bm{Q}^T \bm{I}(\bm{\theta}) \bm{Q})^{-1} \bm{Q}^T
\end{equation}
if $\bm{Q}^T \bm{I}(\bm{\theta}) \bm{Q}$ is invertible.

\subsubsection{CRB for the variance of coordinate estimator}
\label{section:compute CRB}
We are now ready to investigate the CRB for estimating atomic positions from RDC data. Let $\bm{\zeta}= [\bm{\zeta}_1,\ldots,\bm{\zeta}_K]\in\mathbb{R}^{3\times K}$ be the coordinates of the atoms we want to estimate. We aim to derive a lower bound  $\Tr(\bm{Q}(\bm{Q}^T \bm{I}(\bm{\zeta}) \bm{Q})^{-1} \bm{Q}^T)$ of $\mathbb{E}[\Tr((\hat {\bm{\zeta}}-\bm{\zeta})^T(\hat {\bm{\zeta}}-\bm{\zeta}))]$ for any unbiased estimator $\hat{\bm{\zeta}}$ of $\bm{\zeta}$, and compare
\begin{equation}
\sqrt{\frac{\Tr(\bm{Q}(\bm{Q}^T \bm{I}(\bm{\zeta}) \bm{Q})^{-1} \bm{Q}^T)}{K}}
\end{equation}
with the RMSD of the solutions from RDC-SOS and RDC-NOE-SOS in Fig. \ref{figure:CRB compare}.

We assume that the RDC measurements are generated through the noise model in (\ref{RDC noisy}). This noise model is used to get an expression for $\bm{I}(\bm{\theta})$. There are several sets of equality constraints that need to be considered when deriving $\bm{Q}$. We assume that within each rigid unit, the distance between any pair of atoms is fixed. We therefore have a set of equality constraints
\begin{equation}
\label{bond constraint}
d_{nm}^2 = \|\bm{\zeta}_n - \bm{\zeta}_m\|_2^2,\quad (n,m)\in E_\mathrm{fixed}
\end{equation}
where $E_\mathrm{fixed}$ consists of all atom pairs within each and every rigid unit. Without loss of generality, we also consider the constraint
\begin{equation}
\label{centering constraint}
\bm{\zeta}\bm{1} = 0
\end{equation}
which implies the points $\bm{\zeta}_1,\ldots,\bm{\zeta}_K$ are centered at zero. This is due to the fact that
\begin{eqnarray}
\label{lower bound var}
&\ & \Tr((\hat{\bm{\zeta}} - \bm{\zeta})^T(\hat{\bm{\zeta}} - \bm{\zeta})) \cr
&=& \Tr((\hat{\bm{\zeta}}_c - \bm{\zeta}_c - t \bm{1}^T)^T(\hat{\bm{\zeta}}_c - \bm{\zeta}_c - t \bm{1}^T))\cr
&=& \Tr((\hat{\bm{\zeta}}_c - \bm{\zeta}_c)^T(\hat{\bm{\zeta}}_c - \bm{\zeta}_c))  + (1/K)\|t\|_2^2\cr
&\ & -2 \Tr((\hat{\bm{\zeta}}_c - \bm{\zeta}_c)^T t \bm{1}^T )\cr
&=& \Tr((\hat{\bm{\zeta}}_c - \bm{\zeta}_c)^T(\hat{\bm{\zeta}}_c - \bm{\zeta}_c)) + (1/K)\|t\|_2^2\cr
&\geq& \Tr((\hat{\bm{\zeta}}_c - \bm{\zeta}_c)^T(\hat{\bm{\zeta}}_c - \bm{\zeta}_c))
\end{eqnarray}
where $\bm{\zeta}_c$ and $\hat{\bm{\zeta}}_c$ denote the zero centered coordinates and coordinate estimators, and $t$ is the relative translation between $\bm{\zeta}$ and $\hat{\bm{\zeta}}$. Eq. (\ref{lower bound var}) implies that deriving a lower bound for $\mathbb{E}[\Tr((\hat{\bm{\zeta}}_c - \bm{\zeta}_c)^T(\hat{\bm{\zeta}}_c - \bm{\zeta}_c))]$ is sufficient for obtaining a lower bound for $\mathbb{E}[\Tr((\hat {\bm{\zeta}}-\bm{\zeta})^T(\hat {\bm{\zeta}}-\bm{\zeta}))]$. When there are atoms that are constrained to lie on the same plane, we need to add the constraint that any three vectors in the plane span a space with zero volume, i.e.
\begin{equation}
\label{volume constraint}
\det([\bm{\zeta}_i-\bm{\zeta}_j,\bm{\zeta}_k-\bm{\zeta}_l,\bm{\zeta}_m-\bm{\zeta}_n]) = 0
\end{equation}
for atoms $i,j,k,l,m,n$ in the same plane.

We first start with deriving an expression for the Fisher information matrix when RDC data are generated through (\ref{RDC noisy}). From (\ref{RDC noisy}) and (\ref{unit vector}), the likelihood function for the coordinates is
\begin{multline}
p(\{r_{nm}\}_{(n,m)\in E_\mathrm{RDC}} \vert\bm{\zeta}_1,\ldots,\bm{\zeta}_K) =\\
\underset{\substack{(n,m)\in E_\mathrm{RDC}\\j=1,2}}{\mathlarger{\mathlarger{\mathlarger{\Pi}}}} \frac{1}{\sqrt{2\pi \sigma^2}}\\
\exp\bigg(-\frac{\left((\bm{\zeta}_n-\bm{\zeta}_m)^T \bm{S}^{(j)}(\bm{\zeta}_n-\bm{\zeta}_m)-r_{nm}^{(j)} d_{nm}^2\right)^2}{2 d_{nm}^4 \sigma^2}\bigg)
\end{multline}
and the log-likelihood is (up to an additive constant)
\begin{eqnarray}
&&l(\{r_{nm}\}_{(n,m)\in E_\mathrm{RDC}} \vert\bm{\zeta}_1,\ldots,\bm{\zeta}_K)\cr
&:=&\ln p(\{r_{nm}\}_{(n,m)\in E_\mathrm{RDC}} \vert\bm{\zeta}_1,\ldots,\bm{\zeta}_K) \cr
&=& \sum_{\substack{(n,m)\in E_\mathrm{RDC}\\j=1,2}} \frac{-((\bm{\zeta}_n-\bm{\zeta}_m)^T \bm{S}^{(j)}(\bm{\zeta}_n-\bm{\zeta}_m) - r_{nm}^{(j)} d_{nm}^2)^2}{2 d_{nm}^4 \sigma^2}\cr
&=& -\sum_{\substack{(n,m)\in E_\mathrm{RDC}\\j=1,2}} \frac{(\bm{e}_{nm}^T\bm{\zeta}^T \bm{S}^{(j)}\bm{\zeta}\bm{e}_{nm} - r_{nm}^{(j)} d_{nm}^2)^2}{2 d_{nm}^4 \sigma^2}
\end{eqnarray}
where $\bm{e}_{nm} = \bm{e}_n - \bm{e}_m$. The derivative of $l$ with respect to $\mathrm{vec}(\bm{\zeta})$ is then
\begin{multline}
\nabla_{\mathrm{vec}(\bm{\zeta})} l=\\
 -\sum_{\substack{(n,m)\in E_\mathrm{RDC}\\j=1,2}} \frac{2(\bm{e}_{nm}^T\bm{\zeta}^T \bm{S}^{(j)}\bm{\zeta}\bm{e}_{nm} - r_{nm}^{(j)} d_{nm}^2)}{d_{nm}^4 \sigma^2}\\
 (\bm{e}_{nm}\bm{e}_{nm}^T\otimes \bm{S}^{(j)})\mathrm{vec}(\bm{\zeta}).
\end{multline}
It follows from the noise model (\ref{RDC noisy}) and the independence of $\bm{\epsilon}_{nm}^{(j)}$'s that the Fisher information matrix
\begin{multline}
\label{FIM}
\bm{I}(\bm{\zeta})=\mathbb{E}((\nabla_{\mathrm{vec}(\bm{\zeta})} l) (\nabla_{\mathrm{vec}(\bm{\zeta})} l)^T) = \\
4\sum_{\substack{(n,m)\in E_\mathrm{RDC}\\j=1,2}}\\
 \frac{(\bm{e}_{nm}\bm{e}_{nm}^T\otimes \bm{S}^{(j)})\mathrm{vec}(\bm{\zeta})\mathrm{vec}(\bm{\zeta})^T (\bm{e}_{nm}\bm{e}_{nm}^T\otimes \bm{S}^{(j)})}{\sigma^2 d^4_{nm}}
\end{multline}

Having the Fisher information matrix, we now incorporate the constraints in (\ref{bond constraint}) and (\ref{centering constraint}) in order to obtain a bound as in (\ref{constrained CRB}). Stacking the equality constraints (\ref{bond constraint}) into a $\vert E_\mathrm{fixed}\vert \times 1$ matrix, we get
\begin{equation}
f(\mathrm{vec}(\bm{\zeta})) := \begin{bmatrix} \bm{e}_{n m}^T\bm{\zeta}^T \bm{\zeta}\bm{e}_{nm} - d_{n m}^2 \end{bmatrix}_{(n ,m )\in E_\mathrm{fixed}} = 0
\end{equation}
The gradient matrix is thus
\begin{equation}
\bm{Df}(\mathrm{vec}(\bm{\zeta})) = \mathrm{vec}(\bm{\zeta})^T \begin{bmatrix}(\bm{e}_{nm} \bm{e}_{nm}^T \otimes \bm{I}_3)\end{bmatrix}_ {(n ,m )\in E_\mathrm{fixed}}
\end{equation}
where $\bm{Df}(\mathrm{vec}(\bm{\zeta})) \in \mathbb{R}^{\vert E_\mathrm{fixed} \vert \times 3K}$. We note that $\bm{Df}(\mathrm{vec}(\bm{\zeta}))$ is known as the \emph{rigidity matrix} \cite{jackson2007notes}, and the vectors in its null space indicate the direction of infinitesimal motion the atoms can take without violating (\ref{bond constraint}). Even in the case when all pairwise distances between the atoms are known, there is still a 6-dimensional null space for $\bm{Df}(\mathrm{vec}(\bm{\zeta}))$, corresponding to an infinitesimal global rotation and translation to the coordinates $\bm{\zeta}$ that preserves all pairwise distances. We now augment $f(\mathrm{vec}(\bm{\zeta}))=0$ with the centering constraint $\bm{\zeta}\bm{1} = 0$, and this augments $\bm{Df}(\mathrm{vec}(\bm{\zeta}))$ with three rows $\bm{1}^T \otimes \bm{I}_3$, i.e.
\begin{equation}
\label{rigidity matrix}
\bm{Df}(\mathrm{vec}(\bm{\zeta})) = \begin{bmatrix}\mathrm{vec}(\bm{\zeta})^T [(\bm{e}_{nm} \bm{e}_{nm}^T \otimes \bm{I}_3)]_ {(n ,m )\in E_\mathrm{fixed}}\\ \bm{1}^T \otimes \bm{I}_3\end{bmatrix}
\end{equation}
The inclusion of such centering constraint eliminates the three dimensional subspace in the kernel of the rigidity matrix that corresponds to the translational degree of freedom. Let $\bm{Q}$ be an orthonormal basis that spans the null space of $\bm{Df}(\mathrm{vec}(\bm{\zeta}))$. Together with $(\ref{FIM})$ and $(\ref{constrained CRB})$ we obtain the desired CRB. We omit detailing the derivative for constraint (\ref{volume constraint}) but simply note that the inclusion of such constraints eliminates the out of plane infinitesimal motion for atoms lying on rigid planar unit.

\subsubsection{Inclusion of NOE constraints}
We have so far neglected the use of NOE measurements when deriving the CRB. Unlike RDC, the NOE restraints remain more qualitative, with imprecise upper and lower bound \cite{bonvin1996noe} due to the $r^{-6}$ scaling of the interaction. In protein structural calculation, it is customary to include a flat potential well-like penalty (e.g. (\ref{NOE unrelaxed})) in addition to the RDC log-likelihood function derived from RDC, or treat the backbone NOE as inequality constraints on the distances. In any of these cases, when the coordinates $\bm{\zeta}$ strictly satisfy both upper and lower bounds on the distances, the CRB is exactly the same as the CRB derived in Section \ref{section:compute CRB} \cite{gorman1990crb} since the CRB only depends on the local curvature of the log-likelihood function around $\bm{\zeta}$. Therefore when the noise on RDC is large and the NOE restraints are active in determining a coordinate estimator $\bm{\zeta}$, the CRB may no longer serve as a lower bound for the mean squared error of $\hat{\bm{\zeta}}$. In particular, it is possible for $\hat{\bm{\zeta}}$ to have a mean squared error lower than the CRB due to the bias introduced by the NOE (by favoring solutions that satisfies the distance bounds), as observed in Fig. \ref{figure:CRB compare}. A fundamental results in statistical estimation theory-the bias-variance trade-off \cite{wasserman2013all}, states that the mean squared error of an estimator can be obtained from the summation of the variance and squared bias of the estimator. It is possible that with the expense of having some bias, the variance of an estimator can be greatly reduced, resulting a mean squared error that is lower than the CRB \cite[Chapter 7]{wasserman2013all}.

%For the sake of illustrating the improvement one might be able to obtain from additional NOE measurements, we instead consider the following simplistic noise model for the NOE measurements
%\begin{equation}
%\label{dist noise}
%{d_{nm}^\mathrm{NOE}}^2 = \|\bm{\zeta}_n - \bm{\zeta}_m\|^2_2 + \sigma_\mathrm{NOE} \bm{\epsilon}_{nm} = \bm{e}_{nm}^T\bm{\zeta}^T \bm{\zeta} \bm{e}_{nm}+\sigma_\mathrm{NOE} \bm{\epsilon}_{nm},\quad (n,m)\in E_\mathrm{NOE}
%\end{equation}
%where $\bm{\epsilon}_{nm}\sim\mathcal{N}(0,1)$. In the same way we derive the Fisher information matrix in (\ref{FIM}), we now have
%\begin{multline}
%\label{full FIM}
%I(\zeta)=\mathbb{E}((\nabla_{\mathrm{vec}(\zeta)} l) (\nabla_{\mathrm{vec}(\zeta)} l)^T) \\
%=4\sum_{\substack{(n,m)\in E_\mathrm{RDC}\\j=1,2}} \frac{(\bm{e}_{nm}\bm{e}_{nm}^T\otimes \bm{S}^{(j)})\mathrm{vec}(\zeta)\mathrm{vec}(\zeta)^T (\bm{e}_{nm}\bm{e}_{nm}^T\otimes \bm{S}^{(j)})}{\sigma^2 d^4_{nm}} \\
%+\sum_{\substack{(n,m)\in E_\mathrm{NOE}}} \frac{(\bm{e}_{nm}\bm{e}_{nm}^T\otimes \bm{I}_3)\mathrm{vec}(\zeta)\mathrm{vec}(\zeta)^T (\bm{e}_{nm}\bm{e}_{nm}^T\otimes \bm{I}_3)}{\sigma^2_\mathrm{NOE} }.
%\end{multline}
%The inclusion of the equality distance constraints from $E_\mathrm{fixed}$ in the derivation of CRB is done in the same way as the case without NOE restraints. %We are now ready to compare the performance of our algorithm with the CRB.

\subsubsection{Observed Fisher information matrix and protein variability}
\begin{figure}[h!]
  	\centering
\begin{subfigure}[b]{.5\linewidth}\centering\includegraphics[width=1\textwidth]{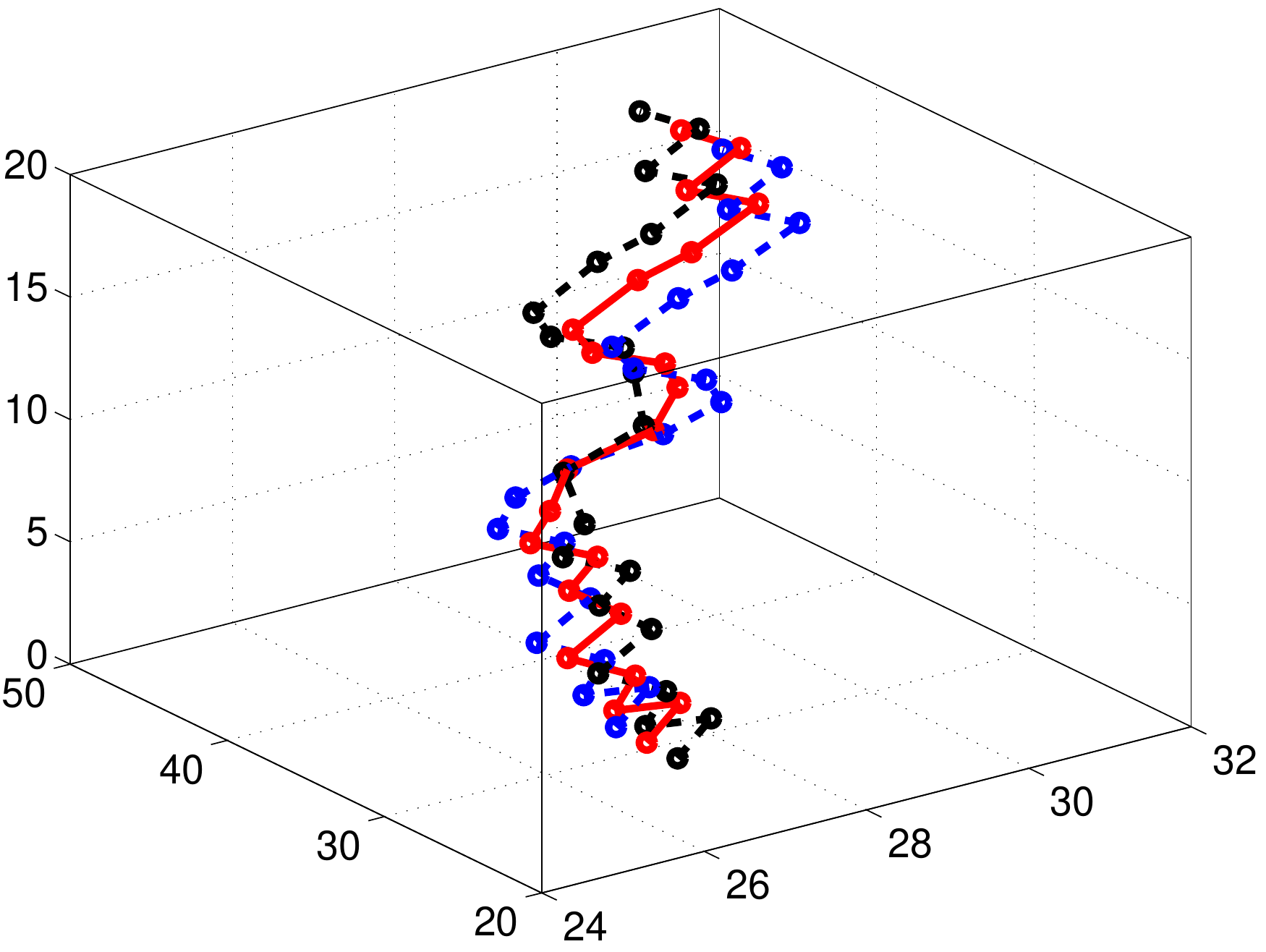}\caption{\centering}\end{subfigure}
\begin{subfigure}[b]{.5\linewidth}\centering\includegraphics[width=1\textwidth]{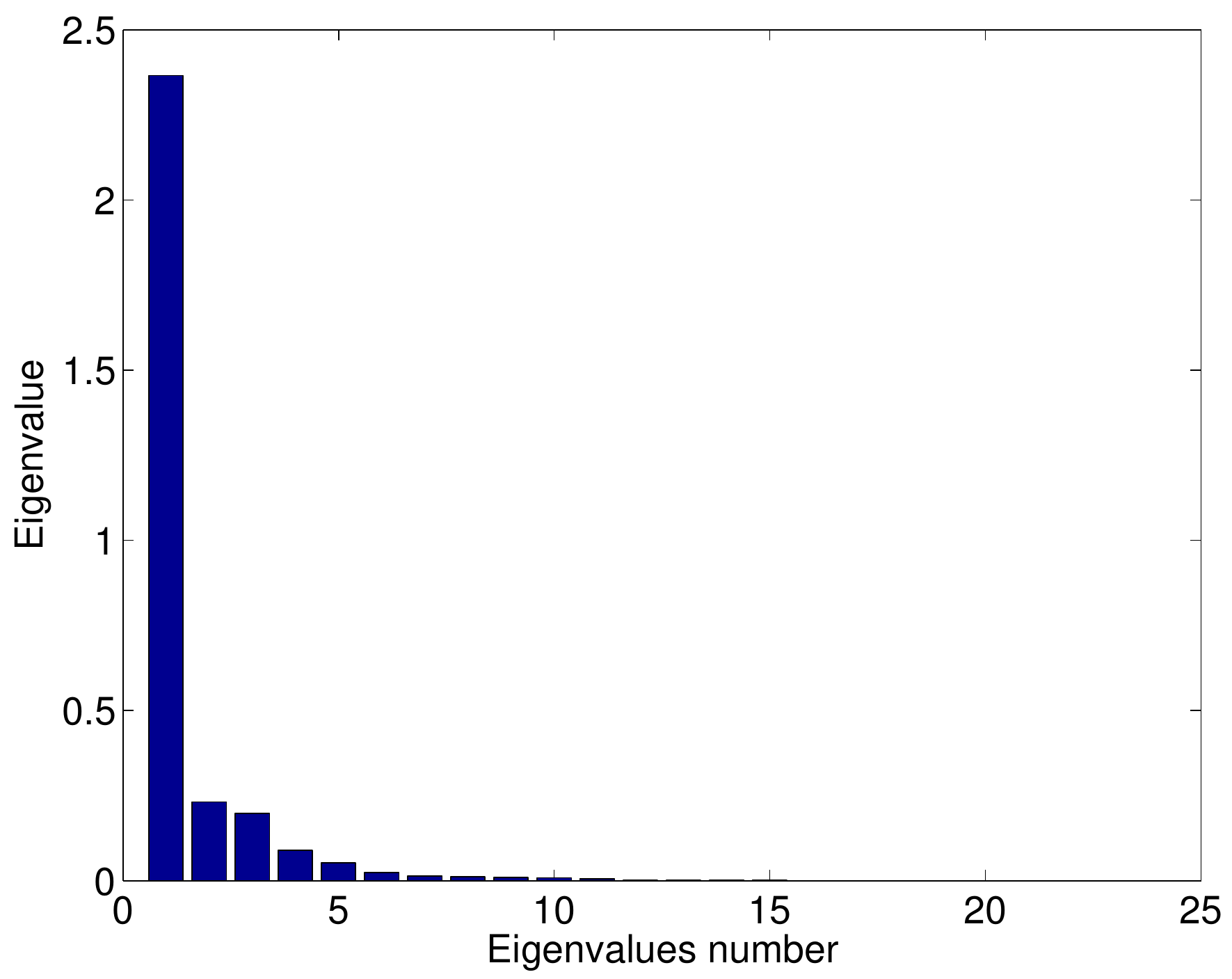}\caption{\centering}\end{subfigure}
\caption{(a)Variation of the structure of a ubiquitin fragment (residue 1-7). We compare the RDC-NOE-SOS solution $\hat{\bm{\zeta}}$ (solid) with two other structures (dashed) obtained from adding the top eigenvector of $\hat{\bm{Q}} (\hat{\bm{Q}}^T \widehat{\bm{I}(\bm{\theta})}^{-1} \hat{\bm{Q}}) \hat{\bm{Q}}^T$ multiplied by small scalars to $\hat{\bm{\zeta}}$. (b) The largest 10 eigenvalues of the inverse of FIM.}\label{figure:variation}
\end{figure}

We remark that since the LHS of (\ref{CRB}) (or (\ref{constrained CRB})) is the covariance matrix for the estimator $\hat{\bm{\theta}}$, the leading eigenvectors of ${\bm{I}(\bm{\theta})}^{-1}$ give the direction of the greatest variations of the protein based on the observed data, whereas the corresponding eigenvalues give the variance (amplitude) of variations. When deriving the CRB, we use the FIM (\ref{FIM}) which is obtained from averaging over the distribution of the data. An estimator $\widehat{\bm{I}(\bm{\theta})}$ of the FIM can be obtained from the observed data, by replacing $\bm{\theta}$ in FIM by its maximum-likelihood estimator $\hat{\bm{\theta}}$ and plugging in the observed data (in our case the observed RDC $r_{nm}$) instead of taking expectation over the distribution of the data. The direction for which an estimator $\hat{\bm{\theta}}$ has the greatest variance can be estimated by the top eigenvector of $\widehat{\bm{I}(\bm{\theta})}^{-1}$. In the constrained case, we compute the top eigenvector of $\hat{\bm{Q}} (\hat{\bm{Q}}^T \widehat{\bm{I}(\bm{\theta})}^{-1} \hat{\bm{Q}}) \hat{\bm{Q}}^T$ where $\hat{\bm{Q}}$ are computed based on $\hat{\bm{\theta}}$ instead of $\bm{\theta}$. In Fig. \ref{figure:variation}a, we demonstrate the variation of the ubiquitin fragment for residue 1-7 (with 159 atoms) using the eigenvector of estimated FIM. In Fig. \ref{figure:variation}b we show the largest 10 eigenvalues of the inverse of the estimated FIM. As we see, there is one prominent mode of variation for this protein fragment. We note that this procedure of determining the modes of protein variation bear resemblance to normal mode analysis \cite{case1994normal}. In such analysis, the Hessian for the pseudo energy function of a protein near a local minimum is first determined. Then the normal modes are determined by the eigenvectors of the Hessian matrix. If we treat the log-likehood function as some pseudo energy function, our FIM-based analysis of the modes of atomic displacement corresponds to the classical normal modes analysis.

\subsubsection{Infinitesimal rigidity and invertibility of the Fisher information matrix}
In this subsection, we study the infinitesimal rigidity \cite{liberti2014euclidean} of the protein structure given RDC and distance measurements and how it guarantees invertibility of the Fisher information matrix. Let a framework with coordinates $\bm{\zeta}\in \mathbb{R}^{3\times K}$ be constrained by
\begin{eqnarray}
\label{framework constraint 1}
(\bm{\zeta}_n - \bm{\zeta}_m)^T (\bm{\zeta}_n - \bm{\zeta}_m) = d_{nm}^2,\quad (n,m)\in E_\text{fixed},\cr
%(\bm{\zeta}_n - \bm{\zeta}_m)^T (\bm{\zeta}_n - \bm{\zeta}_m) = d_{nm}^2,\quad (n,m)\in E_\text{NOE},\cr
\end{eqnarray}
and
\begin{gather}
\label{framework constraint 2}
(\bm{\zeta}_n - \bm{\zeta}_m)^T \bm{S}^{(j)}(\bm{\zeta}_n - \bm{\zeta}_m) = r^{(j)}_{nm},\cr
j=1,\ldots,N,\ (n,m)\in E_\text{RDC}.
\end{gather}
In order to derive a condition for infinitesimal rigidity, we first let $\text{vec}(\bm{\zeta}(s))$ be a curve in dimension $\mathbb{R}^{3K}$ parameterized by $s$, where $\bm{\zeta}(0)$ satisfies (\ref{framework constraint 1}) and (\ref{framework constraint 2}). Taking derivative of the constraints in (\ref{framework constraint 1}) and (\ref{framework constraint 2}) with respect to $s$ at $s=0$, we have

\begin{multline}
\left[\begin{smallmatrix}\mathrm{vec}(\bm{\zeta}(0))^T[\bm{e}_{nm} \bm{e}_{nm}^T \otimes \bm{I}_3]_{(n ,m )\in E_\text{fixed}}\\
%\mathrm{vec}(\bm{\zeta}(0))^T[\bm{e}_{nm} \bm{e}_{nm}^T \otimes \bm{I}_3]_{(n ,m )\in E_\text{NOE}}\\
\mathrm{vec}(\bm{\zeta}(0))^T[\bm{e}_{nm} \bm{e}_{nm}^T \otimes \bm{S}^{(j)}]_{(n ,m )\in E_\text{RDC}, j\in[1,N]}\end{smallmatrix}\right] \frac{d}{ds} \text{vec} (\bm{\zeta}(0))\\
 = \bm{R}(\bm{\zeta}(0))  \frac{d}{ds} \text{vec} (\bm{\zeta}(0)) = 0.
\end{multline}
The null space of the generalized rigidity matrix $\bm{R}(\bm{\zeta}(0))$ with dimension $(\vert E_\text{fixed} \vert +  \vert E_\text{RDC} \vert )\times 3K$ represents the direction of infinitesimal motion such that $\bm{\zeta}(s)$ satisfies the constraints (\ref{framework constraint 1}), (\ref{framework constraint 2}) for infinitesimally small $s$. If $\bm{R}(\bm{\zeta}(0))$ only has a three dimensional nullspace, i.e. the global translations in $x,y,z$-directions, we say the framework $\bm{\zeta}(0)$ along with the constraints (\ref{framework constraint 1}) and (\ref{framework constraint 2}) is infinitesimally rigid.

Now we verify that the constrained Fisher information matrix is invertible if $\bm{R}(\bm{\zeta}(0))$ has a three dimensional null space corresponds to global translation of the points. We define $\text{ker}(\bm{A})$ to be the kernel of a matrix $\bm{A}$ and $\text{range}(\bm{A})$ to be the column space of $\bm{A}$. Let $\bm{Q}$ again be the basis of the nullspace of $\bm{Df}(\mathrm{vec}(\bm{\zeta}))$ defined in (\ref{rigidity matrix}) such that $\bm{Df}(\mathrm{vec}(\bm{\zeta}))\bm{Q} = 0$.  Let $\bm{v}$ satisfies $$ \bm{Q}^T \bm{I}(\bm{\zeta}) \bm{Q} \bm{v} = 0$$  $ \bm{Q}^T  \bm{I}(\bm{\zeta}) \bm{Q} \bm{v} = 0$ if and only if $\bm{v}\in \text{ker}(\bm{Q})$ or $\bm{Q}\bm{v}\in \text{ker}(\bm{I})$. Since the columns of $\bm{Q}$ are linearly independent, $\bm{Q}\bm{v}\neq 0$ unless $\bm{v}=0$. This means $ \bm{Q}^T  \bm{I}(\bm{\zeta}) \bm{Q} \bm{v} = 0$ if and only if $\bm{v} = 0$ or $\bm{Q}\bm{v} \in \text{ker}(\bm{I})\cap \text{range}(\bm{Q}) = \text{ker}(\bm{I})\cap \text{range}(\bm{Q}) = \text{ker}(\bm{I})\cap \text{ker}(\bm{Df}(\mathrm{vec}(\bm{\zeta})))$. Therefore if $$\text{ker}(\bm{I})\cap \text{ker}(\bm{Df}(\mathrm{vec}(\bm{\zeta}))) = \emptyset,$$ or in other words
\begin{equation}
\label{range condition}
\text{range}(\bm{I}) \cup \text{range}(\bm{Df}(\mathrm{vec}(\bm{\zeta})))=\mathbb{R}^{3K}
\end{equation}
then $\bm{Q}^T \bm{I}(\bm{\zeta}) \bm{Q}$ is invertible. From the form of the (\ref{FIM}), it is easy to show that the range condition (\ref{range condition}) is satisfied if and only if the range of
\begin{multline}
\left[\begin{smallmatrix}
\bm{1}^T \otimes \bm{I}_3\\
\mathrm{vec}(\bm{\zeta}(0))^T[\bm{e}_{nm} \bm{e}_{nm}^T \otimes \bm{I}_3]_{(n ,m )\in E_\text{fixed}}\\
%\mathrm{vec}(\bm{\zeta}(0))^T[\bm{e}_{nm} \bm{e}_{nm}^T \otimes \bm{I}_3]_{(n ,m )\in E_\text{NOE}}\\
\mathrm{vec}(\bm{\zeta}(0))^T[\bm{e}_{nm} \bm{e}_{nm}^T \otimes \bm{S}^{(j)}]_{(n ,m )\in E_\text{RDC}, j\in[1,N]}\end{smallmatrix}\right] = \left[\begin{smallmatrix} \bm{1}^T \otimes \bm{I}_3\\ \bm{R}(\bm{\zeta}(0)) \end{smallmatrix}\right]
\end{multline}
is $\mathbb{R}^{3K}$. Then we arrive at the conclusion that if the framework $\bm{\bm{\zeta}}$ is infinitesimally rigid with the null space of $\bm{R}(\bm{\zeta})$ being the global translations, the constrained Fisher information matrix defined as $\bm{Q}^T \bm{I}(\bm{\zeta}) \bm{Q}$ is invertible.

In \cite{yershova2011rdc}, it is shown that if there exists RDC measurements for a bond in the peptide plane and a bond in the CA-body in a single alignment media, the solutions of the protein structure form a discrete set. Therefore under this condition, there is no infinitesimal motion other than global translation such that the protein framework satisfies the RDC and NOE constraints. We can thus compute the CRB safely under such condition.

\bibliographystyle{amsplain}
\bibliography{bibref}
% BibTeX users please use one of
%\bibliographystyle{spbasic}      % basic style, author-year citations
%\bibliographystyle{spmpsci}      % mathematics and physical sciences
%\bibliographystyle{spphys}       % APS-like style for physics
%\bibliography{}   % name your BibTeX data base

\end{document}